\documentclass[10pt]{article}
\usepackage{makeidx}
\usepackage{multirow}
\usepackage{multicol}
\usepackage[dvipsnames,svgnames,table]{xcolor}
\usepackage{graphicx}
\usepackage{epstopdf}
\usepackage{ulem}
\usepackage{hyperref}
\usepackage{amsmath}
\usepackage{amssymb}
\author{-}
\title{}
\usepackage[paperwidth=612pt,paperheight=792pt,top=50pt,right=46pt,bottom=50pt,left=46pt]{geometry}

\makeatletter
	{\par\setlength{\parindent}{#3}
	\setlength{\leftmargin}{#1}       \setlength{\rightmargin}{#2}%
	\advance\linewidth -\leftmargin       \advance\linewidth -\rightmargin%
	\advance\@totalleftmargin\leftmargin  \@setpar{{\@@par}}%
	\parshape 1\@totalleftmargin \linewidth\ignorespaces}{\par}%
\makeatother 

\begin{document}

{\Huge Volumetric segmentation of muscle compartments using in vivo imaging and
architectural validation in human finger flexors }

Yang Li

\textbf{\textit{Abstract}---Segmenting muscle compartments and measuring their
architecture can facilitate movement function assessment, accurate
musculoskeletal modeling, and synergy-based electromyogram simulation. Here, we
presented a novel method for volumetric segmentation of muscle compartments using
in vivo imaging, focusing on the independent compartments for finger control of
flexor digitorum superficialis (FDS). Besides, we measured the architectural
properties of FDS compartments and validated the segmentation. Specifically,
ultrasound and magnetic resonance imaging (MRI) from 10 healthy subjects were
used for segmentation and measurement, while electromyography was utilized for
validation. A two-step piecewise segmentation was proposed, first annotating
compartment regions in the cross-sectional ultrasound image based on compartment
movement, and then performing minimum energy matching to register the ultrasound
data to the three-dimensional MRI coordinate system. Additionally, the
architectural properties were measured in the compartment masks from the
segmentation using MRI tractography. Anatomical correctness was verified by
comparing known anatomy with reconstructed fiber tracts and measured properties,
while segmentation accuracy was quantified as the percentage of finger
electromyogram centers falling within their corresponding compartments. Results
demonstrated agreement for the fiber orientation between the tractography and
cadaveric photographs. Significant differences in architectural properties (fiber
length and physiological cross-sectional area, \textit{P} $<$ 0.001) were
observed between compartments. The properties of FDS and its compartments were
within the physiological ranges (\textit{P} $<$ 0.01). 95\% (38/40) of the
electromyogram centers were located within respective compartments, with 2 errors
occurring in the index and little fingers. The validated segmentation method and derived architectural properties may advance biomedical applications.}

\label{PointTmp}\textbf{\textit{Index Terms}---finger flexor, muscle
compartment, segmentation, anatomy, architectural measurement.}

{\footnotesize Yang Li is with the State Key Laboratory of Mechanical System and
Vibration, School of Mechanical Engineering, Shanghai Jiao Tong University,
China. This preprint does not provide a complete list of authors and will be updated in the future.}

\section{INTRODUCTION}

{\Huge H\label{OLE_LINK4}}and grip is essential for activities of daily living. The function is primarily controlled by the finger flexor muscles, particularly the flexor digitorum
superficialis (FDS). FDS comprises four regions, termed muscle compartments [4],
which can be partially activated during finger flexion to separately control the
index, middle, ring, and little fingers. The localization of muscle compartments
is formally termed muscle compartment segmentation (MCS). MCS and subsequent
architectural measurements are critical for biomedical applications. For
instance, compartment-specific architectural data are clinically utilized to
assess finger function and diagnose muscular disorders [5]. The compartment data
are also crucial for accurate musculoskeletal modeling, which~integrates
musculoskeletal data to study muscle dynamics. However, current approaches
estimate architectural parameters of muscle compartments through indirectly
scaling [6, 7] rather than measurement, which may introduce significant
deviations of force output. Another example is Electromyogram modeling, which
uses subject-specific muscle architecture to study the electrical outputs
generated by skeletal muscles [8]. Muscle fibers, the representatives of muscle
architecture, are distributed and activated independently in groups within
distinct FDS compartments [9], according to the~motor neuron synergy theory.
Thus, characterizing muscle compartment architecture is essential for
electromyogram modeling [10] and neural decoding [11] consistent with the synergy
theory. Extracting the electrical outputs of specific functional regions will also contribute to the development of electromyographic interfaces.

The research of in vivo MCS has mainly utilized magnetic resonance imaging (MRI)
techniques in current practice. Jeneson et al. [12] identified anatomical
compartments in human finger flexors using T2-weighted MRI following grip
exercise. Hojo et al. [13] visualized compartment boundaries in finger flexors
during vibrational excitation via MRI elastography. These studies leveraged
post-exercise metabolites and real-time mechanical vibration, respectively as
compartment markers, thereby selectively activating finger-specific functional
compartments [12]. However, MCS of individual finger remains a major challenge,
given that the research-grade MRI protocols have only achieved preliminary and
unverified segmentation on a single cross-section. Critically, no existing method
enables volumetric MCS, precluding measurement of compartment-specific
architecture essential for quantitative biomedical applications. Some studies
attempted to describe the distribution of muscle compartments from high-amplitude
regions in electromyogram (muscle activation maps) [11, 14]. However, the
multi-compartment architecture led to overlapping regions that have not been
verified anatomically in electromyogram. Furthermore, architectural parameters
cannot be obtained from electromyography.

These limitations stem from the architecture-function of finger flexors and the
challenges of imaging technique. Muscle compartments exhibit a mapping to finger
movements. Compartment fasciae may be sub-millimeter in size or absent [15].
These characteristics necessitate tracking movement-dependent markers rather than
direct anatomical markers as the principle for MCS. Given this principle,
specialized MRI protocols appear to be necessary. However, such protocols
typically incur narrow fields of view and high costs in terms of both time and
personnel [13]. Therefore, relying exclusively on MRI may not constitute an
optimal solution for volumetric MCS. Combining MRI and ultrasound imaging may
provide an in vivo imaging alternative. Compared to static 3D imaging of MRI,
ultrasound enables lower-cost dynamic 2D imaging [16], which shows potential for
capturing compartment movements during finger movements. However, volumetric MCS
requires transformation of multiple 2D images into a 3D volume, i.e., unification
of acquired images into a common coordinate system. The challenge can be overcome
by image registration, where data from different imaging sources are aligned and
combined into a coordinate system [17]. Volumetric MCS can be achieved by
registering 2D ultrasound images to 3D anatomical references of MRI.

Following segmentation, architectural parameters of muscle compartments need to
be calculated to provide references for biomedical applications. Reliable method
for the measurement [18] has been established based on MRI tractography,
including: 1) ``Fiber tractography'': In diffusion-weighted imaging, continuous
fiber tracts are reconstructed from locally estimated discrete fiber orientations
[19]. 2) ``Calculation of architectural parameters'': Architectural parameters of
muscles or compartments are derived from these reconstructed fiber tracts.

In addition, validation of in vivo segmentation is important, as the credibility
of both the MCS and architectural measurement would otherwise be compromised.
However, the unavailability of in vivo actual values hinders the quantitative
validation. Note that the repeatability of measurement [20] does not demonstrate
the validity. Comparative analysis with anatomical references [21] serves as a
validation method for both reconstructed fiber tracts and measured parameters,
which may indirectly indicate the validity of segmentation. Furthermore, we
incorporated surface electromyogram to validate the segmentation. High-density
surface electromyogram can delineate an unbiased estimate of the two-dimensional
longitudinal muscle activation map corresponding to individual finger movement
[11], thereby identifying the dominant muscle compartment activated. The
activation map can then be compared with the compartment region derived from the
segmentation.

This study aims to achieve volumetric MCS of the finger flexor FDS using in vivo
imaging, including MRI and ultrasound imaging. Furthermore, based on the
segmentation, it aims to measure the architecture of muscle compartments and
validate the segmentation. The contributions are summarized as follows:

\textbf{1) "A novel method for volumetric MCS":} We proposed a method called
piecewise segmentation for MCS within muscle volume. The method aligned and
combined 2D ultrasound images into a 3D anatomical MRI coordinate system to
generate volumetric MCS. Piecewise segmentation consists of two components: a)
``Annotation'', using ultrasonography and estimation algorithm for compartment
movements. b) ``Registration'', utilizing minimum energy matching for the fusion
of two image modalities. It is necessary to ensure the compartment specificity
for annotation, i.e., to ensure that the movement is not affected by the
co-contraction between fingers. We~hypothesize that finger co-contraction will be
significantly reduced, when combining voluntary movement of the target finger
[22] with passive constraint on non-target fingers [23]. Both movement
trajectories captured by data gloves and tissue direction fields derived from
ultrasound images indicated predominant activation of the target finger.

\textbf{2) "In vivo architectural properties of FDS muscle compartments":} We
reported in vivo architectural properties of muscle compartments in FDS, based on
data collected from 10 healthy subjects. Following volumetric MCS, the
architectural measurement [18] was conducted in two stages. First, fiber
tractography was performed to generate uniformly distributed fiber tracts within
each muscle compartment. Second, compartment architectural parameters were
calculated based on the fiber tracts. Additionally, we measured the architecture
of FDS, accounting for volumetric constraints across all compartments. The
distribution of these architectural properties exhibited consistency with
anatomical reference.

\textbf{3) "Validation of muscle compartment architecture":} We validated the
architecture of muscle compartments by comparing known anatomical properties with
reconstructed fiber tracts and measured parameters. Additionally, we proposed a
novel validation method for piecewise segmentation according to the consistency
of architecture-function. Specifically, the segmentation was validated by
demonstrating the consistency between the regions of a) muscle compartment from
the segmentation and b) muscle activation from the electromyogram. This
``two-source validation'' was built upon the hypothesis: It would be extremely
unlikely for the two methods to make the exact same mistake [24]. 95\% of the
electromyogram centers (38 in 40) were located within the boundaries of their
respective muscle compartments.

\section{Methods}

The architectural measurement in this study primarily reference our previous
work [18], where muscle architecture was measured using MRI. Accordingly, Section
``MRI Protocol and Preprocessing'', and ``Architectural Measurement'' followed
the methods.

\subsection{Study Population}

The research protocol adhered to the principles outlined in the Declaration of
Helsinki and was approved by the institutional ethics review board (No. E2021178C
and E20240248I). All subjects completed consent forms. Participants were informed
to maintain normal activity levels but avoid intense workouts for one day
preceding data acquisition. Among 12 medically qualified right-handed volunteers
who underwent right forearm imaging, complete datasets from 10 individuals (8
men, 2 women; age = 25$\pm{}$3 years; BMI = 24$\pm{}$3 kg/m${^2}$) were retained
for analysis, with 2 excluded due to motion artifacts. Ultrasound and
electromyogram recordings were additionally acquired from these 10 participants.

\subsection{MRI Protocol and Preprocessing }

MRI was executed on the 3T scanner (MAGNETOM Prisma, Siemens Healthcare,
Erlangen, Germany) featuring the gradient hardware (80 mT/m gradient strength,
200 T/m/s slew rate) coupled with an 18-element flexible torso coil array.
Subjects were in the supine orientation with their right forearm maintained in a
relaxed, neutral position, and with a custom-engineered device designed to
minimize motion artifacts and prevent muscular compression (\textbf{Fig. 1}).

Two imaging sequences were implemented, i.e., T2-weighted images were for
anatomical reference of piecewise segmentation and diffusion-weighted images were
for fiber tractography in architectural measurement.

1) T2-weighted images: Utilizing simultaneous multi-slice accelerated T2 turbo
spin-echo sequence (field of view (FOV) = 120$\times{}$120$\times{}$282 mm${^3}$,
matrix = 448$\times{}$448, slices = 94, resolution =
0.27$\times{}$0.27$\times{}$3 mm${^3}$, repetition time/echo time (TR/TE) =
3500/39 ms, total scan duration = 3 m 36 s).

2) Diffusion-weighted images: Employing a readout-segmented RESOLVE diffusion
tensor sequence (FOV = 102$\times{}$130$\times{}$260 mm$^{3}$, matrix =
80$\times{}$160, slices = 34, resolution = 3$\times{}$1.6$\times{}$1.6 mm${^3}$,
TR/TE = 5500/69 ms, b-value = 400 s/mm${^2}$ with 12 diffusion encoding
directions plus 4 interspersed b = 0 references, with spectral fat suppression,
parallel imaging using 2-times accelerated GRAPPA, total scan duration = 4 m 52
s).

Signal-to-noise ratio [25] was 40$\pm{}$9 for anatomical images and 26$\pm{}$3
for b = 0 images, and it was similar to the baseline [18]. The diffusion imaging
parameters, including b-value selection, directional encoding, and achieved
signal-to-noise ratio, satisfied established criteria for reliable skeletal
muscle diffusion tensor analysis [26]. Representative axial anatomical images
with corresponding diffusion tensor images were provided in \textbf{Supplementary
Material 1}.

MRI processing involved distortion correction for eddy currents, motion
artifacts, and susceptibility effects in the DTI data through FSL eddy 6.0 [27],
utilizing~\textit{b}=0 field maps as reference. Subsequent noise reduction was
achieved via a localized PCA filtering approach [28]. Piecewise registration [29]
was performed to align DTI data to T2-weighted images for anatomical constraints
from T2-weighted images. Piecewise registration involved: 1) initial axial
alignment guided by superficial markers [30] (vitamin D capsules around the wrist
and elbow), followed by 2) subdivision into \textasciitilde{}15-slice segments
for individualized affine transformation. Diffusion gradient orientations were
adjusted through B-matrix rotation [31].

Manual segmentation of FDS muscle was conducted in ImageJ [32] based on the
T2-weighted anatomical images, producing muscle masks that underwent cubic spline
interpolation across slices. In addition to being used as regions of interest for
tractography and MCS, the masks were adopted as the spatial reference for both
ultrasound and electromyogram acquisition (\textbf{Supplementary Material 2}).
The location for acquisition was determined by calculating relative distances to
anatomical landmarks at the wrist and elbow.

\subsection{Ultrasound Imaging Protocol and Preprocessing}

B-mode ultrasound imaging was performed using the Vantage-256 (Verasonics, USA).
A 128-channel linear array probe (L11-5v, center frequency of 7.7 MHz) was
employed. Plane wave imaging was utilized, with an imaging frame rate set at 500
Hz. The imaging field covered both a longitudinal depth and a lateral width of
40.0 mm in the cross-section. The longitudinal and lateral resolutions are
approximately 0.2 mm and 0.3 mm respectively.

Subjects were seated with their upper limbs placed in a custom-designed platform
and the probe fixed at the designated location on the forearm (\textbf{Fig. 2}).
Similar to the segmentation of FDS, images were acquired along the longitudinal
axis of the forearm at about 3 cm intervals (\textbf{Supplementary Material 2}).
Fingers except the target finger were immobilized. Subjects were instructed to
perform reciprocating flexion movements of the target finger for 10 seconds with
the assistance of a metronome at 60 bpm, and to minimize involuntary movements of
the other fingers. To verify the dominant movement of the target finger, movement
angles of the proximal interphalangeal joint were synchronously collected using a
data glove. The angles were normalized based on the maximum angle.

During ultrasonic data acquisition, raw ultrasound radio frequency signals were
recorded using the Verasonics platform. Subsequently, in-phase and quadrature
(IQ) demodulation was applied to the radio frequency data to extract the
amplitude and phase information, resulting in IQ data. Finally, a delay-and-sum
beamforming algorithm was used to reconstruct high-quality B-mode ultrasound
images from the IQ data. The imaging and preprocessing ensured high
temporal--spatial resolution and data quality [33].

The preprocessing, subsequent segmentation and architectural analyses, were
executed using custom Python-based utilities [18].

\subsection{Piecewise Segmentation}

This study proposes a novel method named "piecewise segmentation", for in vivo
volumetric MCS. The core hypothesis for identifying muscle compartments is that,
when finger movement is primarily restricted to the target finger, the movement
direction of its corresponding compartment differs from the other compartments
[34]. Notably, movement amplitude was not selected as there are active and
passive components during muscle contraction [35], which means other compartments
may exhibit comparable passive movement amplitude. Prior to the segmentation, FDS
contours were generated by manual annotation in both ultrasound and MRI images
using ImageJ [32]. Subsequent segmentation was constrained within FDS contours
serving as the region of interest. As illustrated in \textbf{Fig. 3}, piecewise
segmentation consists of two sequential stages: 1) Annotation: Compartment
contours were delineated on individual muscle cross-sections using ultrasound
imaging and movement estimation. 2) Registration: Multiple 2D ultrasound-derived
contours were aligned into a 3D anatomical MRI coordinate system.

\textbf{1) Annotation.} First, Farneback optical flow algorithm [36] was applied
to ultrasound images of the target finger to estimate tissue direction field
(\textbf{Fig. 3a)}). This direction field revealed differences in the movement
direction between muscle compartments (\textbf{Fig. 4(a)}). Farneback method
operated under the assumption that the brightness of a pixel remained unchanged
between frames.

\begin{equation}
{f_2}(x) = {f_1}(x + {\rm{d}})
\label{eq:1}
\end{equation}

$f(x)$: brightness at pixel $x$, and $d$: displacement, represented by
amplitude and direction.~The displacement was computed by fitting the brightness
equation (details in [36]). Farneback optical flow robustly estimates the
movement direction at pixels and has been open-sourced in OpenCV.

To highlight the region of target compartment, we proposed an image segmentation
method named region growing algorithm (\textbf{Fig. 3b)}), which integrated the
direction field. Starting from a seed point, the algorithm expanded the region by
incorporating neighboring pixels based on similarity criteria until no further
pixels met the criteria. Formally:

\begin{equation}
{\mathop{\rm If}\nolimits} \;g(x,y) < t \Rightarrow y \in {\rm{X}}
\label{eq:2}
\end{equation}

$g(x,y)$: similarity metric between~pixels, $x$: a pixel in current segmented
region ${\rm{X}}$, $y$: a candidate pixel in the 8-connected neighborhood of $x$,
and $t$: similarity threshold. The similarity metric follows.

\begin{equation}
g(x,y) = \angle ({\rm{m}},{\rm{n}}) < t
\label{eq:3}
\end{equation}

\begin{equation}
t = {a_{\max }}(1 - r) + {a_{\min }}r
\label{eq:4}
\end{equation}

\begin{equation}
r = d(q,y)/\sqrt {A/\pi }
\label{eq:5}
\end{equation}

$m$: mean direction vector of the current segmented region, updated
iteratively during growth, $n$: direction vector at the candidate pixel $y$, $t$:dynamic angular (similarity) threshold, ${a_{\max
}}$=30$^\circ{}$, ${a_{\min }}$=5$^\circ{}$: upper and lower angular bounds, $r$:
angular scaling factor, $d(q,y)$: Euclidean distance of the seed point  a$q$nd
the candidate pixel $y$, and $\sqrt {A/\pi } $: radius of the equivalent circle
of FDS muscle area $A$ in the ultrasonic cross-section. The pseudo-code of region
growing is provided in \textbf{APPENDIX A}. The dynamic angular threshold $t$
increased as $d$ decreased, which prioritized pixel inclusion near the seed
point. This design can be explained by the fact that the seed point was
placed~according to anatomy (\textbf{Fig. 5}) and~the direction
field~(\textbf{Fig. 4(a)}). The seeds were theoretically within the target
compartment, and pixels distant from the seeds were more likely to belong to
non-target compartments, consequently leading to termination of outward growth.
No more than two seed points were placed per compartment, reflecting the
directional coherence within the target compartment. The segmented regions
exhibited distinct spatial distributions across compartments (\textbf{Fig.
4(b)}). Finally, compartment contours were generated via ImageJ [32] according to
the grown regions (\textbf{Fig. 3c)}).

\textbf{2) Registration.} First, farthest point sampling [37] was applied to FDS
muscle in both ultrasound and MRI images to extract uniformly distributed feature
points (\textbf{Fig. 3e)}). The sampling selected a point set with optimal
spatial uniformity in a 2D domain by iteratively choosing the point farthest from
the existing set, until a predefined number of points (10,000 in this study) was
obtained. Mathematically, the next sampling pointwas${s_{\mathop{\rm F}\nolimits}
}$ determined by:

\begin{equation}
{s_{\mathop{\rm F}\nolimits} } = \arg \max (\mathop {\min }\limits_{p \in
{\rm{P}}\backslash {\rm{S}}} d(p,s))
\label{eq:6}
\end{equation}

${\rm{S}}$: sampled point set, ${\rm{P}}$: set of all points, ${\rm{P}}\backslash {\rm{S}}$:
unsampled point set, and $d(p,s)$: Euclidean
distance between points $p$ and $s$. The farthest strategy ensures uniformity by
maximizing the minimum distance between successive points. The strategy can be
interpreted geometrically: each subsequent point is at the center of the largest
vacant region.

FDS contours served as a~bridge between ultrasound and MRI,~because the contours
can be delineated from both images based on anatomical fascial boundaries. The
uniformly distributed points in the contours should maintain a spatial one-to-one
correspondence between the two images. From a topological perspective, this
correspondence stems from the topological invariance of muscle architecture under
continuous deformation. To establish this correspondence,~bipartite graph
matching~[38] was adopted to match the point pairs located in each cross section
between the two images (\textbf{Fig. 3f)}).

\begin{equation}
{{\rm{M}}_{\mathop{\rm W}\nolimits} } = \arg \min \sum\limits_{(u \in {\rm{U}},v
\in {\rm{V}})} {{d^2}} (u,v)
\label{eq:7}
\end{equation}

${{\rm{M}}_{\mathop{\rm W}\nolimits} }$: Minimum weight matching (an approach of
bipartite graph matching), formulated as point pairs (10,000 from farthest point
sampling), and ${d^2}(u,v)$: squared Euclidean distance of sampled points
respectively from ultrasound image ${\rm{U}}$ and MRI image ${\rm{V}}$. Bipartite
graph matching can directly use the coordinates of the two images. The
transformation of any coordinate system does not affect the matching due to
topological invariance. The matching ensured that the sum of squared distances
between corresponding points is minimized among all possible matching
configurations. According to the topological invariance of muscle architecture,
the matching minimized muscular deformation distances to determine the spatial
correspondence between the two images, preserve minimum deformation energy and
spatial continuity [17]. Based on this matching result, compartment contours from
ultrasound images can be mapped to the MRI coordinate system (\textbf{Fig. 3g)}).
Specifically, contour mapping (\textbf{APPENDIX B}) located the sampling points
closest to the contour in the ultrasound image and then transformed the points to
the MRI image based on the matching relationship. Finally, volumetric compartment
masks were generated by applying cubic spline interpolation across all contours
at different cross-sections [39] (\textbf{Fig. 3h)}).

A movie in \textbf{Supplementary Material 3 }documented the ultrasound data
acquisition, showcasing finger movement alongside the corresponding ultrasound
imaging. The compartment contours were overlaid onto the movie. Compartment masks
were presented to illustrate the relative positions between compartments, and the
masks were then compared with cadaveric illustrations (\textbf{Fig. 5}).

\subsection{Architectural Measurement}

\textbf{1) ``Fiber tractography''.} Both the masks and DTI data were spatially
normalized by resampling to 1 mm isotropic resolution [40]. Diffusion tensors
were estimated in DSI Studio 2023 [41], followed by deterministic tractography
(step size = 1 mm). Streamlines were terminated if fractional anisotropy fell
below 0.1, curvature exceeded 20$^\circ{}$, or they exited the mask boundary
[20]. Post-tracking, short streamlines ($<$ 10 mm) were filtered out, while
retained ones were smoothed using 3rd-order polynomial fitting for physiological
curvature [42]. To ensure architectural reliability, 3,000 streamlines [43] were
required for each measurement.

Muscle fibers are typically represented using fiber tracts or streamlines, with
the two terms being interchangeable outside density-based quantification [44].
The distribution of streamlines is estimated based on the assumption of uniform
fiber density in biological tissues. To address sampling bias in the streamline
distribution, farthest streamline sampling [18] was implemented. The sampling
method minimizes preferential sampling of clustered or long streamlines to ensure
their uniform distribution. Initially, 10,000 candidate streamlines were
generated from seed points within the mask. A spatially balanced subset of 3,000
streamlines was then remained guided by the distance index between streamlines
(details in [18]).

\textbf{2) ``Calculation of architectural parameters''.} Six architectural
parameters were computed from the masks and the reconstructed fiber tracts,
including:~muscle volume (MV),~fiber length (FL),~muscle length (ML), FL/ML
ratio,~pennation angle (PA), and~physiological cross-sectional area (PCSA).
\label{OLE_LINK17}These architectural properties hold important functional
significance in anatomical and musculoskeletal modeling research [45].

1. MV~was calculated by counting the voxels within the muscle mask and
multiplying by the voxel volume (1 mm${^3}$).

2. FL~for a streamline was determined by summing the distances between
consecutive points along the streamline.

3. PA~for a streamline was defined as the angle between the streamline and the
line of action. The overall~FL~and~PA~for a muscle (or compartment) were taken as
the median values across all streamlines, given predominantly non-Gaussian
distributions of these properties [43].

4. The~line of action~was derived from the average streamline direction,
and~ML~was measured as the distance between the most proximal and distal points
along this line. Although the line of action can be determined more accurate
based on architectural types of muscles [46], this study did not address the
architectural types, as it falls beyond the scope of this work.

5. FL/ML ratio, an intrinsic property (independent of muscle size), was
calculated for comparison with anatomical studies.

6. PCSA~was computed as described in [47]:

\begin{equation}
{\rm{PCSA}} = {\rm{MV}}*{\rm{cos}}({\rm{PA}})/{\rm{FL}}
\label{eq:8}
\end{equation}

Four architectural parameters of muscle compartments were measured except ML and
FL/ML ratio. Volume fractions of index\&little and middle\&ring are intrinsic
properties, and were calculated as the percentage of the volume to the total
volume of all compartments. The muscle compartment architecture was compared with
anatomical results [2]. Furthermore, the six architectural parameters of FDS were
measured for segmented and non-segmented FDS architecture, by using all
compartment masks and only FDS muscle mask respectively, as termination criteria
for tractography.~A comparative analysis was performed between segmented,
non-segmented, and anatomical [1] FDS architecture. Theoretically, segmented
muscle architecture should better approximate anatomy, as the compartment masks
correctly defined the anatomical constraints of muscle fibers.

\subsection{Cadaveric Photographs}

To validate the fiber orientation of muscle compartments derived from piecewise
segmentation, we compared the fiber tractography with cadaveric muscle bellies
from about 80-year-old donors [1, 2]. The cadaver specimens had been preserved in
formalin and showed no signs of prior surgical intervention. The dissection
protocol involved: 1) fixation in a neutral position, 2) removal of skin and
subcutaneous tissue, and 3) meticulous dissection of connective tissue overlying
each finger muscle (or muscle compartment).

\subsection{Surface Electromyogram}

Participants were seated with the forearm naturally positioned on a
custom-designed platform, and with high-density surface electromyography to
record muscle activity of FDS (\textbf{Fig. 6}). Skin preparation included
alcohol swab cleansing prior to electrode placement. The electrode configuration
consisted of two grids with 64 electrodes per grid (5$\times{}$13 arrangement
with an inter-electrode distance of 8 mm in two directions). At the designated
location on the forearm (S\textbf{upplementary Material 2 }and F\textbf{ig. 6)},
two electrode grids were placed on the muscle belly parallel to the muscle's
longitudinal axis. Prior to the experiment, maximum voluntary contraction (MVC)
of isometric flexion was measured for each finger excluding the thumb. The target
finger and the force gauge were connected through a finger stall. The other
fingers were restricted in their natural state and were required to minimize
involuntary contraction. Subsequently, three trials of single-finger isometric
flexion at 30\% MVC were performed, each lasting 12 seconds.

\hspace{15pt}Electromyogram signals were recorded in monopolar mode with
band-pass filtering (10-4400 Hz), sampled at 2048 Hz with a 150-gain using a
multichannel acquisition system (EMG-Quattrocento, OT Bioelettronica, Italy). The
acquired signals underwent 20-500 Hz band-pass filtering (4th-order Butterworth)
followed by 50 Hz comb filtering to eliminate power-line interference [14].
Initial and final 1-second segments were cut to eliminate transient artifacts.
Root mean square (RMS) amplitude was calculated for each channel. Outlier
channels (RMS values beyond mean$\pm{}$3 standard deviation of all channels) were
excluded. Normalized RMS maps as muscle activation maps (0-1 scale, excluding
outliers) were generated by averaging three trials.

The center of surface electromyogram is the weighted average position, which is
used to evaluate the spatial distribution of muscle activity or activation [14].
The centerof $\bar x$the RMS-derived muscle activation map was utilized to locate
the muscle compartment of the target finger.

\begin{equation}
\bar x = \sum\nolimits_{i = 1}^n {({r_i}}  \cdot {x_i})/\sum\nolimits_{i = 1}^n
{{r_i}}
\label{eq:9}
\end{equation}

${x_i}$: coordinate point in the coordinate system of the electrodes, with a total
of $n$ points, and ${r_i}$: RMS value at ${x_i}$. The coordinates with RMS $>$ 0.8
were utilized to calculate the center. Based on the designated location and size
of the electrodes, the voxels of the electrode surface were determined in the
MRI-derived forearm mask. Then, the average normal direction of the electrode
surface was taken as the projection direction, and the boundaries of muscle
compartments from the segmentation were projected onto the electrode surface. The
projected boundaries can reflect the variation of muscle width along its length,
rather than just taking a fixed width in [14]. The electromyogram signals of the
target finger should conduct within the corresponding compartment. Therefore, if
the center falls within the compartment contour, it is considered to have passed
the ``two-source validation'' and achieved architecture-function consistency. The
accuracy was calculated as the number of centers falling within the contour
divided by the total number of centers.

\subsection{Statistical Analysis}

Since the Shapiro-Wilk test confirmed that the architectural data of subjects
followed a normal distribution, parametric tests were employed. Notably, muscle
volume (MV) was excluded from subsequent analysis due to its high inter-subject
variability, which complicates cross-study comparison. Besides, FDS volume was
determined sorely by anatomical images and remained unaffected by piecewise
segmentation. A paired t-test was used to assess the significance of differences
in segmented and non-segmented FDS architectural properties, with Bland-Altman
plots utilized for visualizing these differences. Specifically, scatter points of
the differences were plotted, with linear regression analysis to detect potential
proportional bias (i.e., whether the differences were influenced by the mean
values). When a statistically significant linear correlation was present, the
regression fit line was displayed [48]. Otherwise, only the scatter points were
retained. A one-way analysis of variance was performed to examine the differences
in architectural properties across compartments. Post hoc analysis was conducted
using Bonferroni-corrected paired t-tests. Box plots were employed to compare the
data between this study and anatomical literature. PA of human forearm was less
than 30$^\circ{}$, with smaller values ($<$10$^\circ{}$) observed in the finger
flexors. FL/ML ratio was 0.2--0.6 [49]. A one-sample t-test was used to assess
whether the mean properties of FDS and its compartments fell within these
physiological ranges. Data were showed as mean$\pm{}$standard deviation unless
otherwise specified. The significance threshold was set at $\alpha{}$ = 0.05.
Statistical analysis was performed via IBM SPSS Statistics 25.

\section{Results}

\subsection{Piecewise Segmentation}

The finger co-contraction was significantly reduced, as evidenced by two
observations: 1) the dominant movement of the target finger and 2) the dominant
activation of the target compartment. First, the normalized angular
representation of finger movement trajectories (\textbf{Fig. 7}) demonstrated
that the target finger exhibited the greatest range 0-1 of movement. Second, as
evidenced by the tissue orientation field in ``Methods'' (\textbf{Fig. 4}), the
movement direction of the target compartment differed from that of other
compartments, which was also the basis for identifying muscle compartments.

\textbf{Supplementary Material 3} demonstrated the movement process of all
compartments during the movement of target finger. A directional difference can
be observed between the active movement of the target compartment and the passive
movement of other compartments. This experiment preliminarily validated the
potential of proposed segmentation method for real-time visualization of
compartment movement.

\textbf{Fig. 5} illustrated the relative positions of the muscle compartments.
In \textbf{Fig. 5(a)}, the middle and ring (3rd and 4th fingers) were positioned
more superficially relative to the index and little (2nd and 5th fingers) [3].
The deep compartments were more ulnar than the superficial. When observing the
compartments from the ventral forearm aspect, in the superficial layer
(\textbf{Fig. 5(b)}), 4th finger was positioned on the ulnar side of 3rd finger.
In the deep layer (\textbf{Fig. 5(c)}), 5th finger lied ulnar to 2nd finger.
These relative compartment positions showed agreement between cadaveric
illustrations [3] and compartment masks from the segmentation.

\subsection{Fiber Orientation Validation}

The compartment masks were overlaid onto anatomical images to verify data
alignment, and fiber tractography was displayed in different encoding schemes
(\textbf{Supplementary Material 4}). \textbf{Fig. 8} demonstrated tractography
compared with cadaveric photographs. While the fiber orientation between finger
muscles showed similarity along the forearm axis, key differences [2] were
observed: 1) Index, middle and ring fingers: Proximal fibers aligned along the forearm axis. Distal fibers progressively deviated from radial to ulnar inclination. 2) Little finger: Fibers exhibited an overall oblique orientation from radial-to-ulnar direction. The proximal fibers of the little were oriented at a more oblique orientation compared to those of the other fingers, which resulted in shorter proximal fibers indicated by white dashed lines in \textbf{Fig. 8(b)}. The fiber tractography showed correspondence with cadaveric anatomy.

\subsection{Architectural Properties}

Architectural parameters of FDS and its compartments were measured from 10
subjects (\textbf{Supplementary Table 1}). The architectural properties were within the
physiological ranges in PA of 0-10$^\circ{}$ and FL/ML ratio of 0.2-0.6 (both
\textit{P}$<$0.01), for segmented FDS, non-segmented FDS and the compartments. No
significant differences were observed in PA and ML (\textit{P}=0.953 and 0.125,
respectively). Whereas, the segmentation significantly affected FL, FL/ML ratio,
and PCSA (\textit{P}$<$0.001, 0.001 and 0.01, respectively; \textbf{Fig. 9}).
Furthermore, Bland-Altman analysis revealed no significant linear regression for
differences of FL and FL/ML ratio. Therefore, only scatter plots of the
differences were shown in \textbf{Fig. 10(c)(d)}. The segmentation resulted in a
27\% reduction in FL and a 23\% decrease in FL/ML ratio. For PCSA, a significant
proportional bias was detected (\textit{P}$<$0.05, \textit{R}=0.64), indicating
that the segmentation-induced difference varied with the mean PCSA value
(F\textbf{ig. 10(e))}. The segmentation led to a 29\% increase in PCSA. Segmented
FDS architecture demonstrated higher consistency with anatomical results,
compared to non-segmented architecture (F\textbf{ig. 9(c)(e))}.

For muscle compartment architecture using the segmentation, significant
differences were observed in~FL~and PCSA~(both~\textit{P}$<$0.001), but not in PA
(\textit{P}=0.899; \textbf{Fig. 11(a)}). Middle finger exhibited the longest FL,
followed by ring, index, and little fingers (\textbf{Fig. 11(b)}). Though middle
finger had the largest PCSA, it showed no significant difference from index
finger. No significant PCSA difference existed between ring and little fingers.
The above similarity resulted in two levels of PCSA values for ring\&little and
index\&middle (\textbf{Fig. 11(c)}).\textbf{ }Volume fraction of middle\&ring
compartments~was 1.3 times that of index\&little~(\textbf{Fig. 11(d)}), which was
close to cadaveric findings reporting 1.6 times [1].

\subsection{Architecture-function Validation}

Muscle activation maps, electromyogram centers and projected boundaries of the
muscle compartments were shown in \textbf{Fig. 12}. The centers fell within the
boundaries (100\% accuracy for the ring and middle), except for the index and
little fingers (90\% accuracy for both in \textbf{Supplementary Material 5}). For
all subjects, the confidence regions of the centers and the average boundaries of
the muscle compartments were displayed (\textbf{Fig. 13}), which indicated that
there were overlapping regions between compartments, especially between the
index\&middle, and the ring\&little fingers.

\section{Discussion}

This study first proposed a method named piecewise segmentation, using
ultrasound imaging and magnetic resonance imaging (MRI) to achieve muscle
compartment segmentation (MCS) within the volume of flexor digitorum
superficialis (FDS). Then, the architectural properties of FDS and its muscle
compartments were measured via MRI fiber tractography. The reconstructed fiber
tracts in each compartment were declared anatomically correct to indirectly
validate the segmentation. Finally, the validity of the segmentation was
confirmed by the consistency between the compartment regions identified through
the segmentation and those derived from electromyogram.

\subsection{Piecewise Segmentation}

The principle of MCS is to track specific movement information of the target
compartment, which should not be affected by the co-contraction between fingers.
To ensure the specificity, unnatural experimental conditions were adopted.
Subjects were required to move the target finger [22], and non-target fingers
were constrained by custom devices [23]. The specificity was demonstrated by the
maximum movement angle of the target finger and the different movement direction
of the target compartment. The limitations of the experimental design are the
small number of subjects and the imbalance in gender distribution, due to the
high cost of data acquisition especially in MRI, though lower than similar
studies [12, 13].

The core hypothesis for identifying compartments is the difference in the
movement direction between the target compartment and other compartments, which
was demonstrated through the tissue direction field and supplementary ultrasound
images. This hypothesis is derived from the studies of the movement direction of
muscle fibers. Zhou et al. [34] used the consistency of the direction within the
target region from optical flow to estimate the change in muscle fiber length.
However, they focused on overall muscular changes rather than local information.
Englund et al. [35] demonstrated the local multi-dimensional direction of muscle
fibers, which may be due to the active and passive components of fiber movement.
The active component came from partially recruited muscle fibers in the muscle,
which can be ensured to mainly be in the target compartment through experimental
conditions. The active and passive components were similar in magnitude because
the muscle almost maintained an equal volume during contraction [35]. This
physiological knowledge leads us to propose a segmentation method based on the
movement direction, named region growing. Seed points were placed based on priori
knowledge, including the relative positions and direction field of the
compartments. The direction consistency of the target compartment made the
growing dependent on no more than two seed points. Therefore, the influence from
the randomness of the seed points was reduced by using prior physiological
knowledge. Similar studies [13] conducted MCS on the muscle cross-section and
produced overlapping compartment regions, making it hard to generate
interference-free volumetric compartments. In contrast, region growing generated
spatial distributions without overlap between compartments by controlling seed
points and growth at specific locations. However, the limitation is that region
growing relies on the specific movement of the target compartment, which requires
experimental conditions. Under natural conditions, this specificity may be
replaced by active co-contractions between fingers. Moreover, the segmentation
error resulting from the co-contraction can be reduced rather than eliminated.

To achieve volumetric MCS, compartment contours in the ultrasound images were
registered slice by slice to the MRI. Observable FDS contours served as a bridge
connecting the two image modalities. Uniformly distributed points in the FDS
contours were used as the feature to determine the optimal alignment of the two
images. The feature was designed to prevent the insufficiency of registration
quality caused by sparsity [17]. According to the topological invariance of
muscle architecture, the muscle should have the minimum deformation energy
between the two images [17]. Therefore, the minimized square distance function
was set as the matching criterion to achieve the minimum deformation energy.
Finally, the compartment contour points located in the FDS contours were mapped
to the MRI system to generate the volumetric compartment masks. Additionally,
muscular deformation caused by the forearm position and the compression of
ultrasonic probe [16] were corrected, as the mapped contours can reflect the
neutral position during MRI acquisition. The limitation lies in the
centimeter-level resolution along the forearm axis being restricted by the
ultrasound equipment.

\subsection{Anatomical Correctness}

The fiber orientation and the architectural properties from fiber tractography
can be compared with anatomy to indirectly validate in vivo segmentation. This
correlation with anatomy is referred to as anatomical correctness. The
limitations arising from cadaveric morphology (such as muscle mass reduction and
contraction) and subjectivity [18], as there is no in vivo true value as a "gold
standard". For each muscle compartment, the fiber direction of tractography and
cadaveric photographs were consistent from proximal to distal. These compartments
were not adjacent to each other as depicted in learning materials [50], but
rather exhibited relative positions in the ulnar-radial and depth directions.
Notably, the little finger photograph in this study was from Asians and the
origin was at the medial epicondyle near the elbow [1], which differs from the
origin that may at the middle of the forearm in other races.

The finger ranking of fiber length (FL) and physiological cross-sectional area
(PCSA) in this study was consistent with the anatomical study [2]. FL and PCSA
respectively indicate the muscular capacity of displacement and force generation
[18]. Therefore, it can be inferred that the middle finger had the strongest
capacity, while the weakest was either the middle or the little. The properties
of fingers may represent different functional roles [43] to meet the demands of
complex tasks. Pennation angle (PA) showed no significant difference among the
fingers and was around 5$^\circ{}$, indicating that the fibers in FDS generally
parallel to the forearm axis. However, this did not represent the local fiber
orientation differences between compartments according to the results of
tractography. The volume fraction can be used to distinguish functional roles
[43] and was only used for anatomical validation in this study.\label{OLE_LINK2}

Segmented FDS architecture had higher consistency with anatomy [1]. Significant
differences in segmented and non-segmented FDS architecture were observed in FL,
FL/ML ratio, and PCSA. According to the formula (8), these differences were
computationally related. The reduction in FL led to a decrease in FL/ML ratio and
an increase in PCSA. The differences can be explained by the anatomical
constraints. The compartment masks from the segmentation terminated the
propagation of fibers and reduced FL. However, the results do not support the
introduction of segmentation merely for FDS architectural measurement. First, the
mean differences of known properties were within the range of 20\% to 30\%,
meaning that these differences could be roughly estimated. Second, the cost of
segmentation may be unacceptable if the error is estimable.

\subsection{Architecture-function Validation}

The architecture-function validation was designed, since the electrical signals
generated by finger movements are conducted in the corresponding compartments.
Most electromyogram centers were located within the compartment boundaries. The
error may result from the co-contraction of the wrist flexors on both sides of
FDS, causing additional activation to stabilize the wrist during finger movements
(\textbf{Supplementary Material 5}). The muscle activation maps were similar to
those of other electromyography studies [11, 22, 51], especially regarding the
relative positions of compartments in the ulnar-radial direction. The validation
provided a reference for electromyography and neural decoding studies, where the
spatial properties of electrical signals are difficult to interpret. The
validation proved that both the anatomical structure and the electromyogram had
wide overlap between compartments, and the complex anatomy may pose challenges in
identifying the myoelectric activities of fingers [51]. These activities are
helpful in determining the force and joint angles of prosthetic hands [11]. The
limitation is that the muscle activation maps were affected by the detection
range of the electrodes, because signals deeper than 20 mm are difficult to
capture [14]. The muscle depth of FDS is usually less than 30 mm [14], which
means that the muscle activation maps may be affected.

\subsection{Applications}

The segmentation method and \label{OLE_LINK5}architectural properties of muscle
compartments in this study is promising for biomedical applications.
Subject-specific architectural information can be used in clinical assessment of
finger function [5]. Referring to the average distribution of architectural
properties, similar distribution can be set in universal musculoskeletal models
[45]. \label{OLE_LINK3}It is also feasible to use the data for subject-specific
models. The fiber tracts and segmented regions can be utilized to study the
influence of specific populations and fingers on electromyogram. Muscle fibers in
different compartments can be independently activated through synergy in
electromyogram modeling [9] to enhance the research on neural decoding [11]. Electromyographic interfaces may achieve more robust and accurate motion estimation by extracting the electrical activity of specific muscles or compartments. Further research is needed to explore additional biomedical applications and determine whether the segmentation method can be adapted to other multi-compartment muscles.

\section{Conclusion}

This study presented a novel method for volumetric muscle compartment
segmentation using in vivo imaging, including ultrasound and magnetic resonance
imaging. Moreover, the in vivo architecture of flexor digitorum superficialis and
its compartments was measured based on tractography. Results showed that the
reconstructed fiber tracts and architectural parameters were anatomically
correct. The segmentation validity was confirmed by the consistency between the
compartments and corresponding electromyogram. The above validation enhanced the
credibility of the segmentation and the architectural properties. The
segmentation is not supported if it is merely for measuring the FDS architecture,
considering the trade-off between error and cost. The segmentation method and
architectural properties can be used to enhance musculoskeletal modeling,
electromyogram simulation and clinical assessment of movement function.

\section{Appendix (Pseudo-code)}

\subsection{Region Growing}

\vspace{3pt} \noindent
\begin{tabular}{p{238pt}}
\hline

{\footnotesize \textbf{Algorithm 1:} Region growing}
 \\
\hline

{\footnotesize \textbf{Input:} Region ${\rm{X}}$, seed point $q$, direction
field ${\rm{V}}$}, {\large  }{\footnotesize threshold-related fixed parameters $A$, ${a_{\max }}$, ${a_{\min }}$.}

{\footnotesize \textbf{Output:} Grown region ${\rm{X}} = [q,{x_1},{x_2},...]$.}

\begin{enumerate}
	\item  {\rm{X}} = [q];
    \item {\footnotesize \textbf{while} $\exists y \in {\mathop{\rm neiborhood}\nolimits} (x),$ $x \in
{\rm{X}}$, \textbf{do}}
	\item \quad {\footnotesize ${\rm{m}} \leftarrow {\rm{V}}({\rm{X}})$; // mean direction vector of {\rm{X}}}
	\item \quad{\footnotesize ${\rm{n}} \leftarrow {\rm{V}}(y)$; //direction
vector of {\rm{y}}}
	\item \quad{\footnotesize   $r = d(q,y)/\sqrt {A/\pi } $; //angular scaling factor}
	\item \quad{\footnotesize   $t = {a_{\max }}(1 - r) + {a_{\min }}r$; //dynamic angular
threshold}
	\item \quad{\footnotesize \textbf{if $\angle ({\rm{m}},{\rm{n}}) < t$},\textbf{ then}}
	\item \quad \quad{\footnotesize     append $y$ to ${\rm{X}}$ as $x$; //${\rm{X}} = [q,{x_1},{x_2},...]$}
	\item \quad\textbf{{\footnotesize   end if}}
	\item \textbf{{\footnotesize end while}}
\end{enumerate}
 \\
\hline
\end{tabular}
\vspace{2pt}

\subsection{Contour Mapping}

\vspace{3pt} \noindent
\begin{tabular}{p{238pt}}
\hline

{\footnotesize \textbf{Algorithm 2:} Contour mapping}
 \\
\hline

{\footnotesize \textbf{Input:} Point pairs ${{\rm{M}}_W}$, including sampled
points $u$ and $v$ respectively from two images ${\rm{U}}$ and ${\rm{V}}$, original
contour points ${\rm{X}} = [{x_1},{x_2},...]$.}

{\footnotesize \textbf{Output:} Mapped contour
points $\;{\rm{Y}} = [{y_1},{y_2},...]$.}

\begin{enumerate}
	\item {\footnotesize \textbf for each point ${x_i}\;$ in ${\rm{X}}$,
\textbf{do}}
	\item \quad {\footnotesize   ${u_n} \leftarrow {\mathop{\rm
nearest}\nolimits}({x_i}) $; //nearest point of ${x_i}$}
	\item \quad {\footnotesize  ${v_n} \leftarrow {{\rm{M}}_W}({u_n})$; // matched
point of ${u_n}$}
	\item \quad {\footnotesize   append ${v_n}$ to ${\rm{Y}}$ as ${y_i}$; //${\rm{Y}} = [{y_1},{y_2},...]$}
	\item \textbf{{\footnotesize end for}}
\end{enumerate}
 \\
\hline
\end{tabular}
\vspace{2pt}

\textbf{References}

{\footnotesize [1]\hspace{15pt}K. Matsuzawa\textit{ et al.}, ``The origin
structure of each finger in the flexor digitorum superficialis muscle,''
\textit{Surg. Radiol. Anat.,} vol. 43, no. 1, pp. 3-10, 2021.}

{\footnotesize [2]\hspace{15pt}E. S. Campisi\textit{ et al.}, ``Intramuscular
aponeuroses and fiber bundle morphology of the five bellies of flexor digitorum
superficialis: A three-dimensional modeling study,'' \textit{J. Anat.,} vol. 242,
no. 6, pp. 1003-1011, 2023.}

{\footnotesize [3]\hspace{15pt}S. Erguen\textit{ et al.}, ``Old name, new face:
A systematic analysis of flexor digitorum superficialis muscle with chiasma
antebrachii,'' \textit{Annals of Anatomy-Anatomischer Anzeiger,} vol. 247, 2023.}

{\footnotesize [4]\hspace{15pt}Y. K. Mariappan\textit{ et al.}, ``Vibration
imaging for localization of functional compartments of the extrinsic flexor
muscles of the hand,'' \textit{J. Magn. Reson. Imaging,} vol. 31, no. 6, pp.
1395-1401, 2010.}

{\footnotesize [5]\hspace{15pt}S. R. Ward\textit{ et al.}, ``High stiffness of
human digital flexor tendons is suited for precise finger positional control,''
\textit{J. Neurophysiol.,} vol. 96, no. 5, pp. 2815-2818, 2006.}

{\footnotesize [6]\hspace{15pt}M. Mirakhorlo\textit{ et al.}, ``A
musculoskeletal model of the hand and wrist: Model definition and evaluation,''
\textit{Comput. Methods Biomech. Biomed. Eng.,} vol. 21, no. 9, pp. 548-557,
2018.}

{\footnotesize [7]\hspace{15pt}M. Zheng\textit{ et al.}, ``Isometric
plantarflexion moment prediction based on a compartment-specific HD-sEMG-driven
musculoskeletal model,'' \textit{IEEE Trans. Biomed. Eng.,} vol. 71, no. 8, pp.
2311-2320, 2024.}

{\footnotesize [8]\hspace{15pt}S. Ma\textit{ et al.}, ``Conditional generative
models for simulation of EMG during naturalistic movements,'' \textit{IEEE Trans.
Neural Networks Learn. Syst.}, 2024.}

{\footnotesize [9]\hspace{15pt}S. Madarshahian\textit{ et al.}, ``Synergic
control of a single muscle: The example of flexor digitorum superficialis,''
\textit{Journal of Physiology-London,} vol. 599, no. 4, pp. 1261-1279, 2021.}

{\footnotesize [10]\hspace{15pt}S. Ma\textit{ et al.}, ``Analytical modelling of
surface EMG signals generated by curvilinear fibers with approximate conductivity
tensor,'' \textit{IEEE Trans. Biomed. Eng.,} vol. 69, no. 3, pp. 1052-1062,
2022.}

{\footnotesize [11]\hspace{15pt}R. Roy\textit{ et al.}, ``Concurrent and
continuous prediction of finger kinetics and kinematics via motoneuron
activities,'' \textit{IEEE Trans. Biomed. Eng.,} vol. 70, no. 6, pp. 1911-1920,
2023.}

{\footnotesize [12]\hspace{15pt}Y. K. Mariappan\textit{ et al.}, ``Magnetic
resonance elastography: a review,'' \textit{Clin. Anat.,} vol. 23, no. 5, pp.
497-511, 2010.}

{\footnotesize [13]\hspace{15pt}E. Hojo\textit{ et al.}, ``MR elastography-based
slip interface imaging (SII) for functional assessment of myofascial interfaces:
A feasibility study,'' \textit{Magn. Reson. Med.,} vol. 92, no. 2, pp. 676-687,
2024.}

{\footnotesize [14]\hspace{15pt}M. Xia\textit{ et al.}, ``Extracting individual
muscle drive and activity from high-density surface electromyography signals
based on the center of gravity of motor unit,'' \textit{IEEE Trans. Biomed.
Eng.,} vol. 70, no. 10, pp. 2852-2862, 2023.}

{\footnotesize [15]\hspace{15pt}S. Ortiz-Miguel\textit{ et al.}, ``Compartments
of the antebrachial fascia of the forearm: clinically relevant ultrasound,
anatomical and histological findings,'' \textit{Surg. Radiol. Anat.,} vol. 43,
no. 10, pp. 1569-1579, 2021.}

{\footnotesize [16]\hspace{15pt}J. Verheul, and S.-H. Yeo, ``A hybrid method for
ultrasound-based tracking of skeletal muscle architecture,'' \textit{IEEE Trans.
Biomed. Eng.,} vol. 70, no. 4, pp. 1114-1124, 2023.}

{\footnotesize [17]\hspace{15pt}E. Ferrante, and N. Paragios, ``Slice-to-volume
medical image registration: A survey,'' \textit{Med. Image Anal.,} vol. 39, pp.
101-123, 2017.}

{\footnotesize [18]\hspace{15pt}Y. Li\textit{ et al.}, ``Optimized uniform
sampling and validation of fiber tracts from magnetic resonance tractography for
in vivo architectural measurement of human forearm muscles,'' \textit{IEEE Trans.
Biomed. Eng.,} vol. 71, no. 12, pp. 3370-3382, 2024.}

{\footnotesize [19]\hspace{15pt}D. K. Jones\textit{ et al.}, ``White matter
integrity, fiber count, and other fallacies: The do's and don'ts of diffusion
MRI,'' \textit{Neuroimage,} vol. 73, pp. 239-254, 2013.}

{\footnotesize [20]\hspace{15pt}B. Bolsterlee\textit{ et al.}, ``Reliability and
robustness of muscle architecture measurements obtained using diffusion tensor
imaging with anatomically constrained tractography,'' \textit{J. Biomech.,} vol.
86, pp. 71-78, 2019.}

{\footnotesize [21]\hspace{15pt}M. Froeling\textit{ et al.}, ``Diffusion-tensor
MRI reveals the complex muscle architecture of the human forearm,'' \textit{J.
Magn. Reson. Imaging,} vol. 36, no. 1, pp. 237-248, 2012.}

{\footnotesize [22]\hspace{15pt}Y. Zheng, and X. Hu, ``Concurrent estimation of
finger flexion and extension forces using motoneuron discharge information,''
\textit{IEEE Trans. Biomed. Eng.,} vol. 68, no. 5, pp. 1638-1645, 2021.}

{\footnotesize [23]\hspace{15pt}T. J. Butler\textit{ et al.}, ``Selective
recruitment of single motor units in human flexor digitorum superficialis muscle
during flexion of individual fingers,'' \textit{Journal of Physiology-London,}
vol. 567, no. 1, pp. 301-309, 2005.}

{\footnotesize [24]\hspace{15pt}D. Farina, and A. Holobar, ``Characterization of
human motor units from surface EMG decomposition,'' \textit{Proc. IEEE,} vol.
104, no. 2, pp. 353-373, 2016.}

{\footnotesize [25]\hspace{15pt}O. Dietrich\textit{ et al.}, ``Measurement of
signal-to-noise ratios in MR images: Influence of multichannel coils, parallel
imaging, and reconstruction filters,'' \textit{J. Magn. Reson. Imaging,} vol. 26,
no. 2, pp. 375-385, 2007.}

{\footnotesize [26]\hspace{15pt}M. Froeling\textit{ et al.}, ``DTI of human
skeletal muscle: the effects of diffusion encoding parameters, signal-to-noise
ratio and T-2 on tensor indices and fiber tracts,'' \textit{NMR Biomed.,} vol.
26, no. 11, pp. 1339-1352, 2013.}

{\footnotesize [27]\hspace{15pt}A. Mastropietro\textit{ et al.}, ``Quantitative
comparison of spherical deconvolution approaches to resolve complex fiber
configurations in diffusion MRI: ISRA-based vs L2L0 sparse methods,''
\textit{IEEE Trans. Biomed. Eng.,} vol. 64, no. 12, pp. 2847-2857, 2017.}

{\footnotesize [28]\hspace{15pt}J. V. Manjon\textit{ et al.}, ``Diffusion
weighted image denoising using overcomplete local PCA,'' \textit{PLoS ONE,} vol.
8, no. 9, 2013.}

{\footnotesize [29]\hspace{15pt}Y. Li\textit{ et al.}, "Tracking forearm muscle
fibers from diffusion MRI during dynamic contractions," in \textit{Proc. The 31st
Annual Meeting of ISMRM}, 2022.}

{\footnotesize [30]\hspace{15pt}M. T. Izatt\textit{ et al.}, ``Determining a
reliably visible and inexpensive surface fiducial marker for use in MRI: a
research study in a busy Australian Radiology Department,'' \textit{Bmj Open,}
vol. 9, no. 8, 2019.}

{\footnotesize [31]\hspace{15pt}A. Leemans, and D. K. Jones, ``The B-Matrix must
be rotated when correcting for subject motion in DTI data,'' \textit{Magn. Reson.
Med.,} vol. 61, no. 6, pp. 1336-1349, 2009.}

{\footnotesize [32]\hspace{15pt}C. A. Schneider\textit{ et al.}, ``NIH Image to
ImageJ: 25 years of image analysis,'' \textit{Nat. Meth.,} vol. 9, no. 7, pp.
671-675, 2012.}

{\footnotesize [33]\hspace{15pt}Z. Yin\textit{ et al.}, ``A blind source
separation algorithm for decoding the mechanical spatiotemporal responses of
motor units,'' \textit{Science China-Technological Sciences,} vol. 68, no. 5,
2025.}

{\footnotesize [34]\hspace{15pt}G.-Q. Zhou, and Y.-P. Zheng, ``Automatic
fascicle length estimation on muscle ultrasound images with an
orientation-sensitive segmentation,'' \textit{IEEE Trans. Biomed. Eng.,} vol. 62,
no. 12, pp. 2828-2836, 2015.}

{\footnotesize [35]\hspace{15pt}E. K. Englund\textit{ et al.}, ``Combined
diffusion and strain tensor MRI reveals a heterogeneous, planar pattern of strain
development during isometric muscle contraction,'' \textit{American Journal of
Physiology-Regulatory Integrative and Comparative Physiology,} vol. 300, no. 5,
pp. R1079-R1090, 2011.}

{\footnotesize [36]\hspace{15pt}G. Farneb\"{a}ck, "Two-frame motion estimation
based on polynomial expansion," \textit{Image Analysis, Proceedings}, Lecture
Notes in Computer Science J. Bigun and T. Gustavsson, eds., pp. 363-370, 2003.}

{\footnotesize [37]\hspace{15pt}A. Konstantin\textit{ et al.}, ``Simulation of
motor unit action potential recordings from intramuscular multichannel scanning
electrodes,'' \textit{IEEE Trans. Biomed. Eng.,} vol. 67, no. 7, pp. 2005-2014,
2020.}

{\footnotesize [38]\hspace{15pt}F. Serratosa, ``Computation of graph edit
distance: Reasoning about optimality and speed-up,'' \textit{Image Vision
Comput.,} vol. 40, pp. 38-48, 2015.}

{\footnotesize [39]\hspace{15pt}B. Bolsterlee\textit{ et al.}, ``How does
passive lengthening change the architecture of the human medial gastrocnemius
muscle?,'' \textit{J. Appl. Physiol.,} vol. 122, no. 4, pp. 727-738, 2017.}

{\footnotesize [40]\hspace{15pt}B. Bolsterlee\textit{ et al.},
``Three-dimensional architecture of the whole human soleus muscle in vivo,''
\textit{PeerJ,} vol. 6, 2018.}

{\footnotesize [41]\hspace{15pt}F.-C. Yeh\textit{ et al.}, ``Deterministic
diffusion fiber tracking improved by quantitative anisotropy,'' \textit{PLoS
ONE,} vol. 8, no. 11, 2013.}

{\footnotesize [42]\hspace{15pt}B. M. Damon\textit{ et al.}, ``Polynomial
fitting of DT-MRI fiber tracts allows accurate estimation of muscle architectural
parameters,'' \textit{Magn. Reson. Imaging,} vol. 30, no. 5, pp. 589-600, 2012.}

{\footnotesize [43]\hspace{15pt}J. Charles\textit{ et al.}, ``From fibre to
function: are we accurately representing muscle architecture and performance?,''
\textit{Biol. Rev.,} vol. 97, no. 4, pp. 1640-1676, 2022.}

{\footnotesize [44]\hspace{15pt}G. Girard\textit{ et al.}, ``Towards
quantitative connectivity analysis: Reducing tractography biases,''
\textit{Neuroimage,} vol. 98, pp. 266-278, 2014.}

{\footnotesize [45]\hspace{15pt}J. P. Charles\textit{ et al.},
``Subject-specific muscle properties from diffusion tensor imaging significantly
improve the accuracy of musculoskeletal models,'' \textit{J. Anat.,} vol. 237,
no. 5, pp. 941-959, 2020.}

{\footnotesize [46]\hspace{15pt}D. Lee\textit{ et al.}, ``A three-dimensional
approach to pennation angle estimation for human skeletal muscle,''
\textit{Comput. Methods Biomech. Biomed. Eng.,} vol. 18, no. 13, pp. 1474-1484,
2015.}

{\footnotesize [47]\hspace{15pt}R. L. Lieber, ``Can we just forget about
pennation angle?,'' \textit{J. Biomech.,} vol. 132, 2022.}

{\footnotesize [48]\hspace{15pt}J. Ludbrook, ``Confidence in Altman-Bland plots:
a critical review of the method of differences,'' \textit{Clin. Exp. Pharmacol.
Physiol.,} vol. 37, no. 2, pp. 143-149, 2010.}

{\footnotesize [49]\hspace{15pt}R. L. Lieber, and J. Friden, ``Clinical
significance of skeletal muscle architecture,'' \textit{Clin. Orthop.}, no. 383,
pp. 140-151, 2001.}

{\footnotesize [50]\hspace{15pt}T. Arakawa\textit{ et al.}, ``Dissection,
digitization, and three-dimensional modelling: a high-fidelity anatomical
visualization and imaging technology,'' \textit{Anat. Sci. Int.,} vol. 98, no. 3,
pp. 337-342, 2023.}

{\footnotesize [51]\hspace{15pt}N. van Beek\textit{ et al.}, ``Activity patterns
of extrinsic finger flexors and extensors during movements of instructed and
non-instructed fingers,'' \textit{J. Electromyogr. Kinesiol.,} vol. 38, pp.
187-196, 2018.}

\newpage
\textbf{Figures}

\begin{figure}[h]
    \centering
    \includegraphics[width=0.48\linewidth]{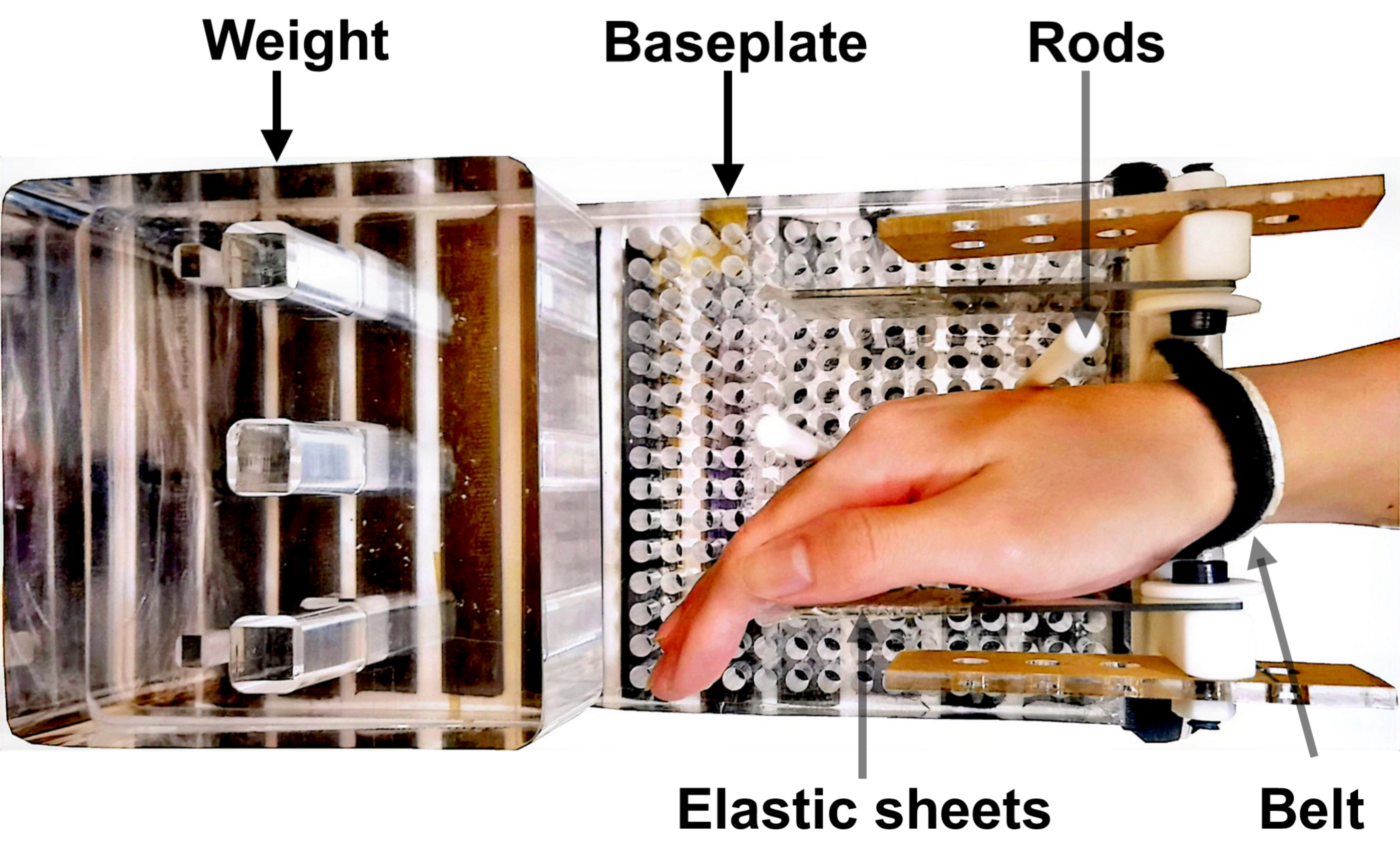}
    \label{fig:1}

\end{figure}

\noindent
\textbf{Fig. 1.}\hspace{15pt}An MRI-compatible hand immobilization device
designed to maintain the forearm in a neutral position. To minimize involuntary
motion during scanning, the baseplate incorporated an array of holes for
inserting plastic rods.

\begin{figure}[h]
    \centering
    \includegraphics[width=0.48\linewidth]{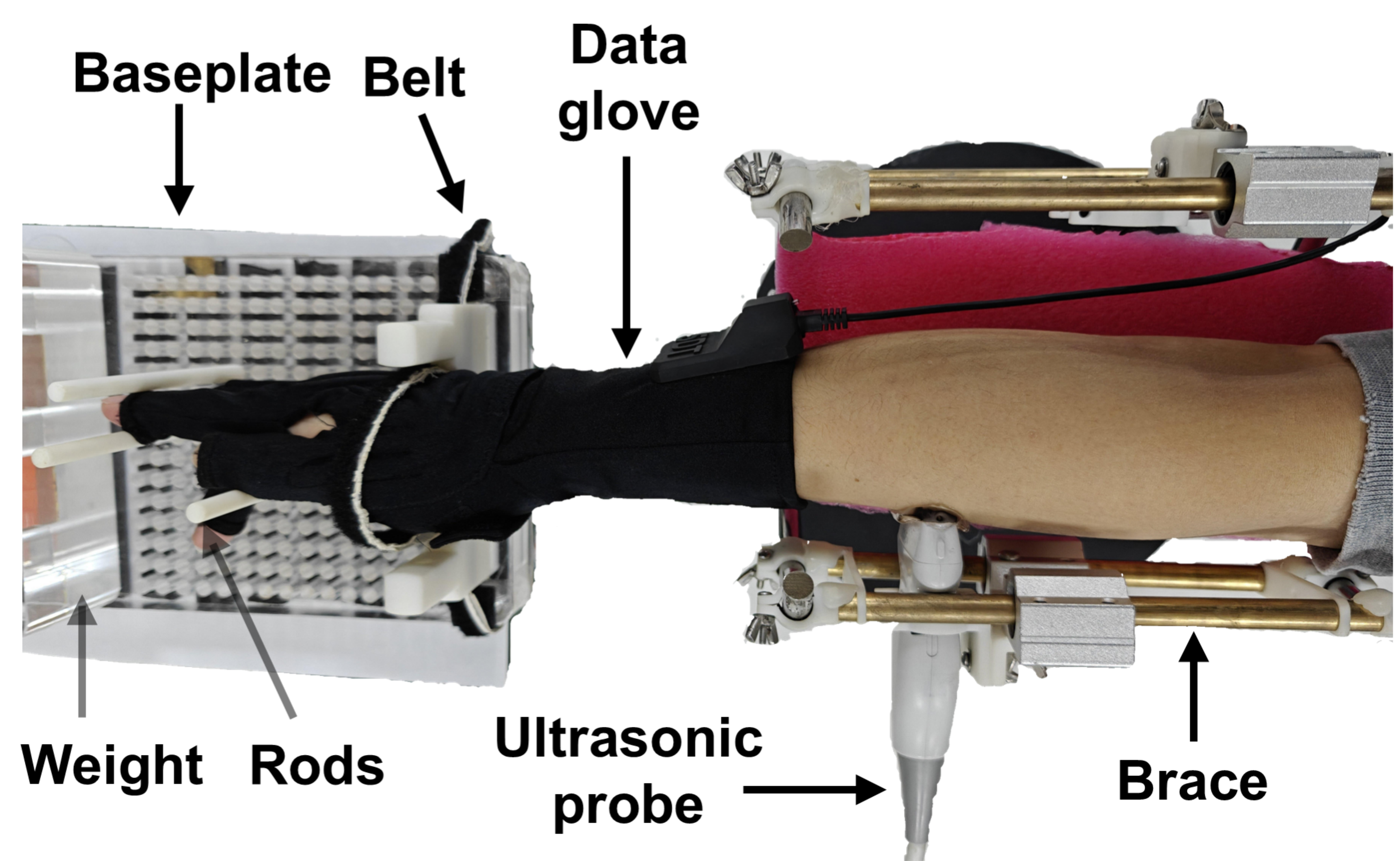}
    \label{fig:2}

\end{figure}

\noindent
\textbf{Fig. 2.}\hspace{15pt}A customized ultrasonic experimental platform
including a hand device and a forearm brace. Both the hand and the forearm were
supported and immobilized. Only the target finger was not restricted to
reciprocating flexion.

\newpage

\begin{figure}[h]
    \centering
    \includegraphics[width=1\linewidth]{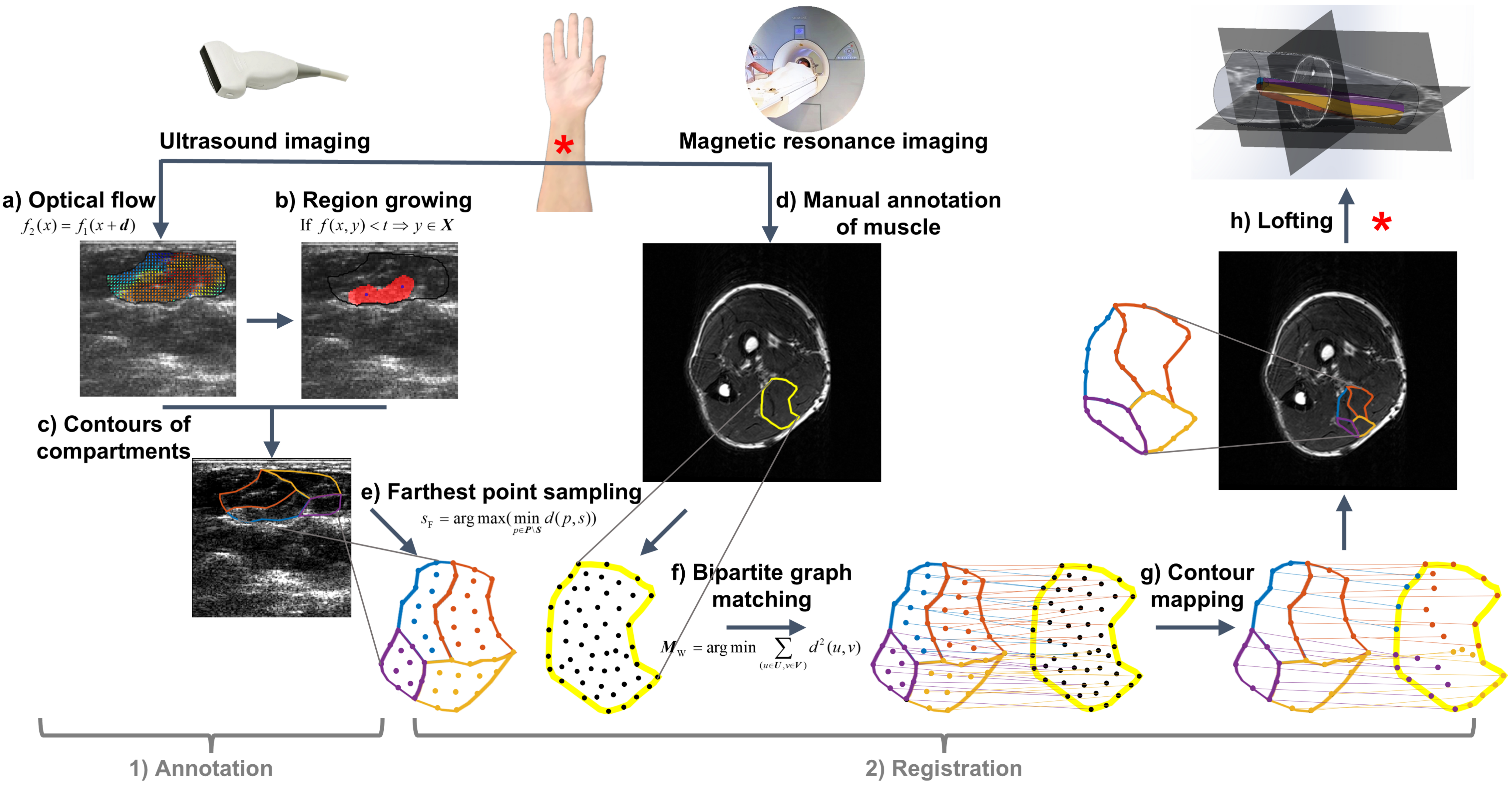}
    \label{fig:3}

\end{figure}

\noindent
\textbf{Fig. 3.}\hspace{15pt}A flowchart of piecewise segmentation, including
annotation and registration. Two red asterisks marked the start and the end,
which were image acquisition and mask generation, respectively. 1) Annotation on
the left consisted of the following: tissue direction field from optical flow,
grown region from region growing, and annotated contour points of muscle
compartments. 2) Registration on the middle and right included the following:
sampled points within FDS contour in both images, bipartite graph matching of the
sampled points, and contour mapping of the contour points. Finally, volumetric
compartment masks were generated by lofting (cubic spline interpolation) of the
contours at different cross-sections.

\begin{figure}[h]
    \centering
    \includegraphics[width=0.9\linewidth]{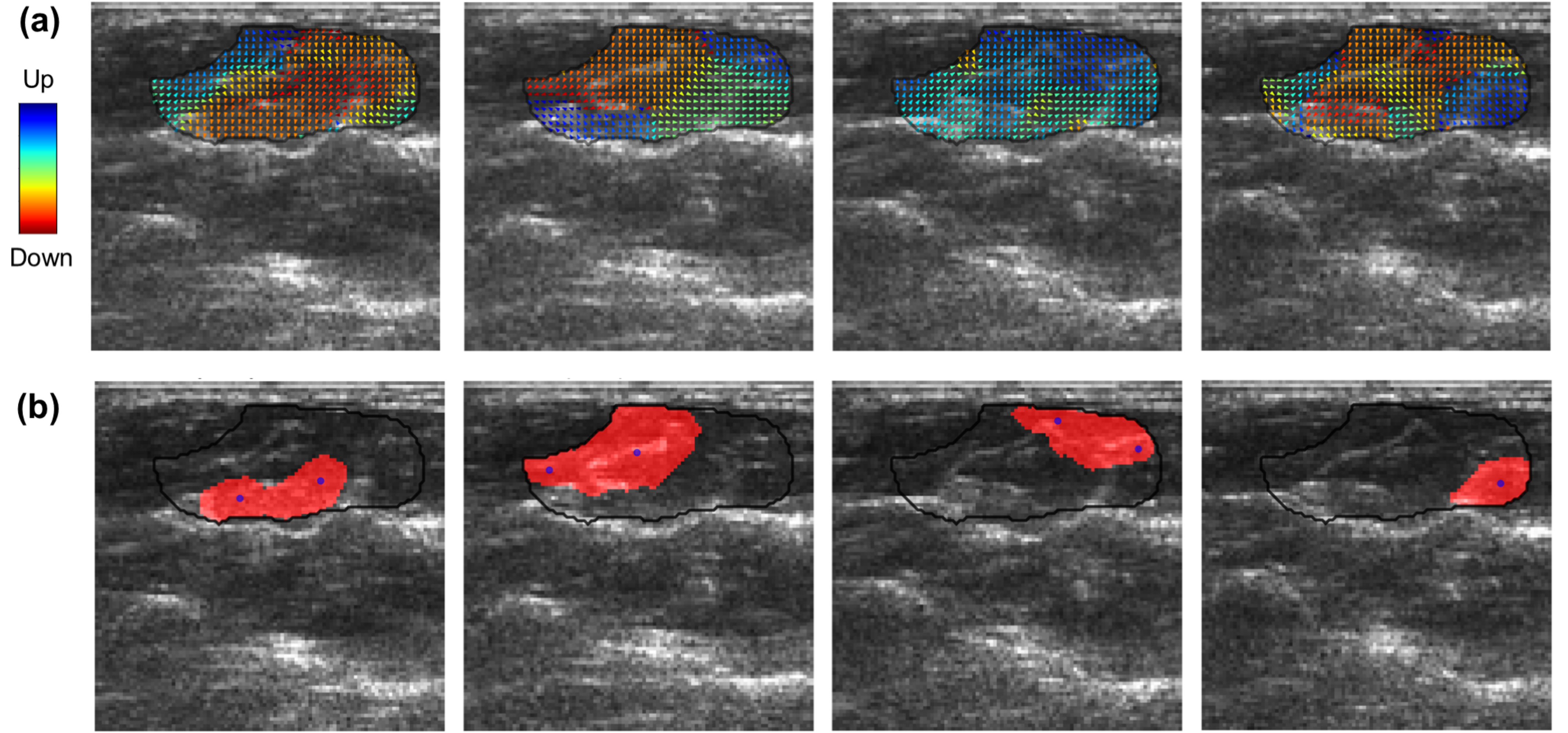}
    \label{fig:4}

\end{figure}

\noindent
\textbf{Fig. 4.}\hspace{15pt}(a) The direction field from Farneback optical flow
and (b) the grown region from region growing. Color coding was used in (a) to
represent movement direction. Blue seed points were marked in the red grown
region of (b). The unique movement direction of the target compartment in (a) can
be observed by combining it with the corresponding grown region in (b).

\newpage

\begin{figure}[h]
    \centering
    \includegraphics[width=0.7\linewidth]{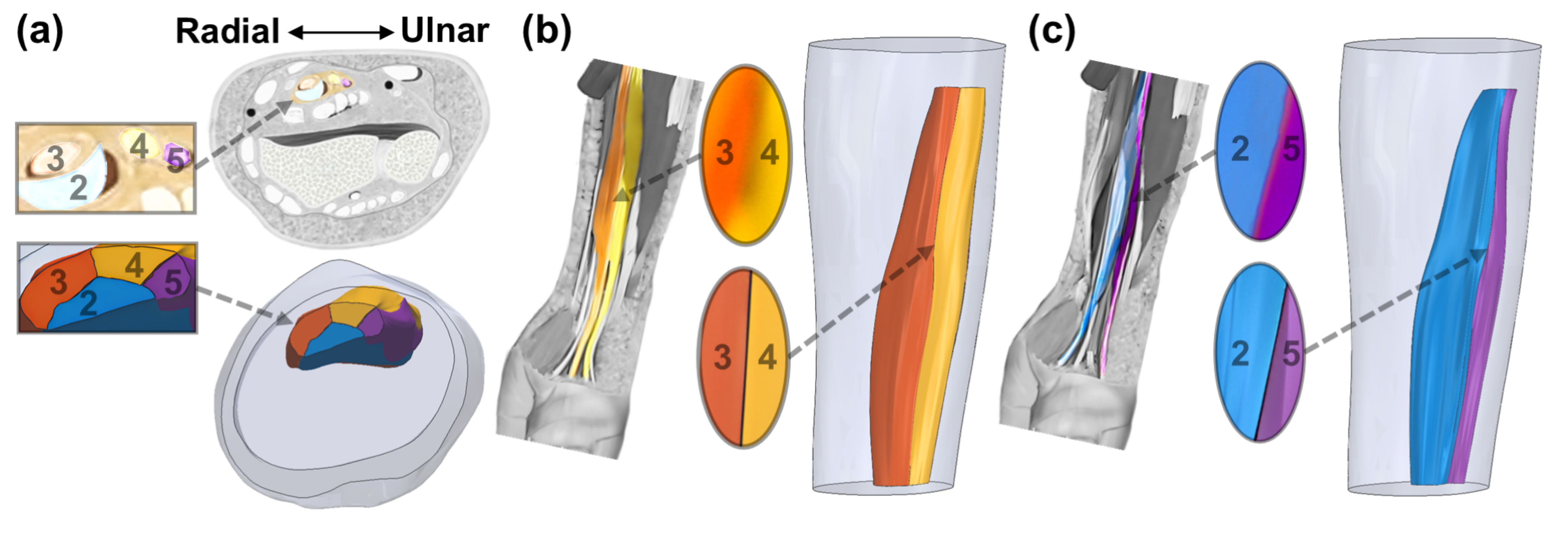}
    \label{fig:5}

\end{figure}

\noindent
\textbf{Fig. 5.}\hspace{15pt}Cadaveric illustrations [3] and compartment masks
from the segmentation, in (a) cross-section, (b) superficial layer and (c) deep
layer on the ventral side of the forearm. (a) showed all four compartments. (b)
displayed the compartments of the middle and ring (3rd and 4th fingers), and (c)
showed the compartments of the index and little (2nd and 5th fingers).

\begin{figure}[h]
    \centering
    \includegraphics[width=0.5\linewidth]{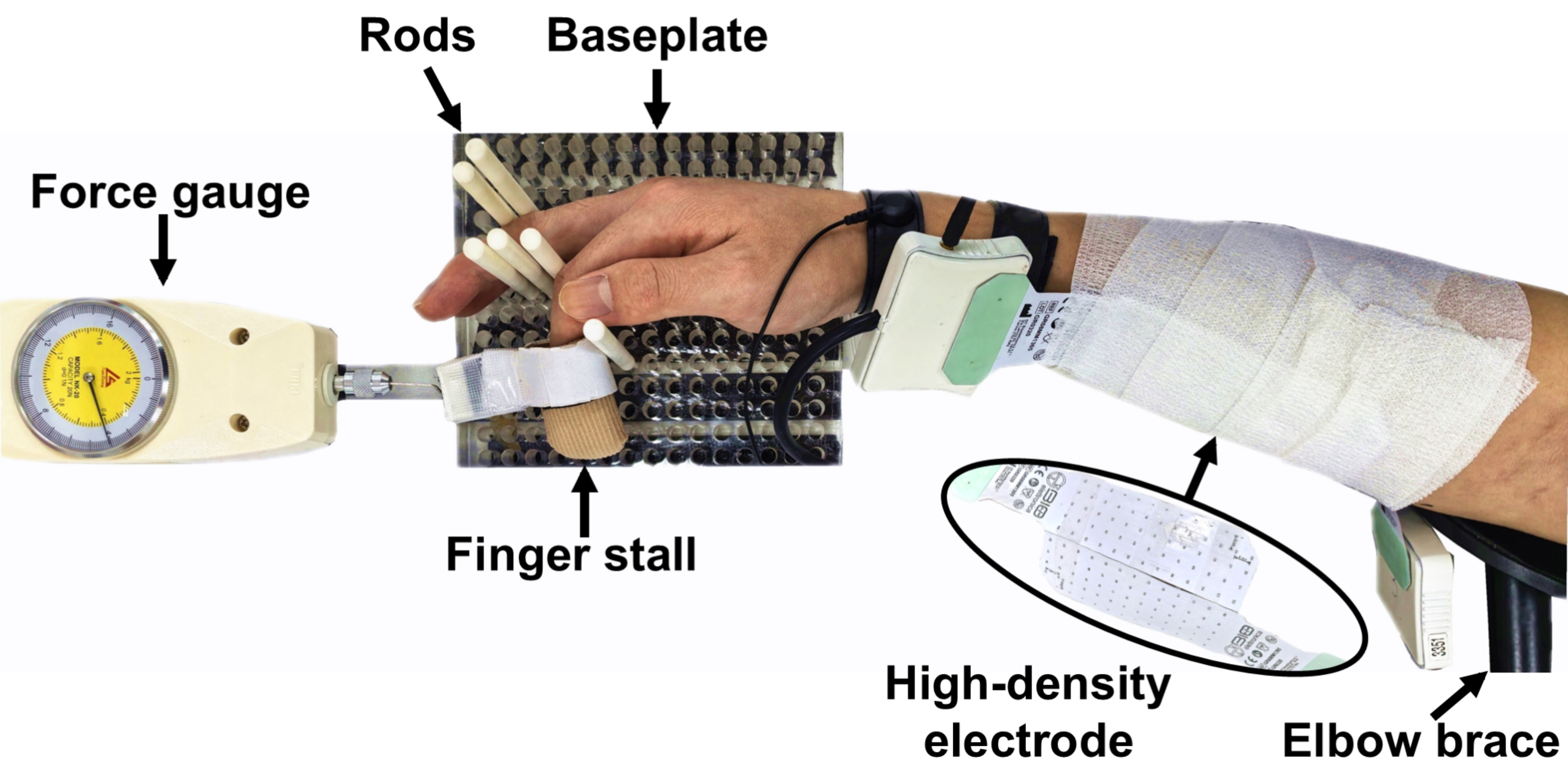}
    \label{fig:6}

\end{figure}

\noindent
\textbf{Fig. 6.}\hspace{15pt}A customized electromyography experimental platform
including a hand device and an elbow brace. Both the hand and the elbow were
supported. Only the target finger performed isometric contractions in a flexed
state, while high-density electromyogram was collected from FDS muscle in the
forearm.

\begin{figure}[h]
    \centering
    \includegraphics[width=0.8\linewidth]{fig_7}
    \label{fig:7}

\end{figure}

\noindent
\textbf{Fi}g. 7.\hspace{15pt}Movement angles of the proximal interphalangeal
joint during reciprocating flexion of the target finger. In the four subplots,
the target finger had the maximum angle, which was used for data normalization.
The movement angles of the other fingers did not exceed approximately 0.3.

\newpage

\begin{figure}[h]
    \centering
    \includegraphics[width=0.7\linewidth]{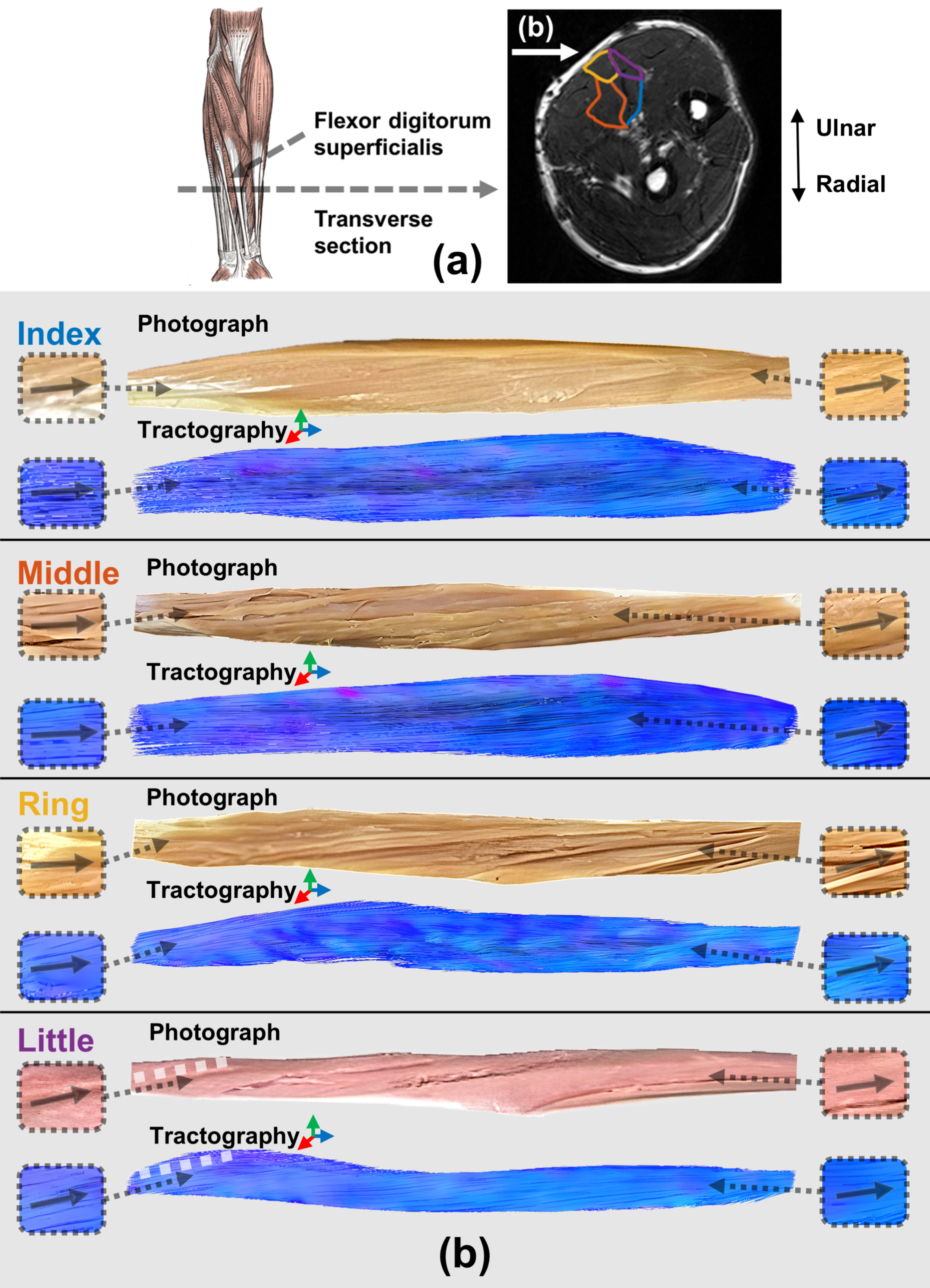}
    \label{fig:8}

\end{figure}

\noindent
\textbf{Fig. 8.}\hspace{15pt}For the muscle compartments of flexor digitorum
superficialis (FDS) in the forearm, (a) T2-weighted magnetic resonance image in
the cross-section, and (b) comparison of tractography and cadaveric photographs
[1, 2] (with permission) in ventral view. In (b), direction-encoded tractography
was employed. Both tractography and photographs were scaled at proximal and
distal regions to display local fiber orientation, indicated by arrows. The short proximal fibers of the little finger were marked with white dotted lines.

\newpage
\begin{figure}[h]
    \centering
    \includegraphics[width=0.63\linewidth]{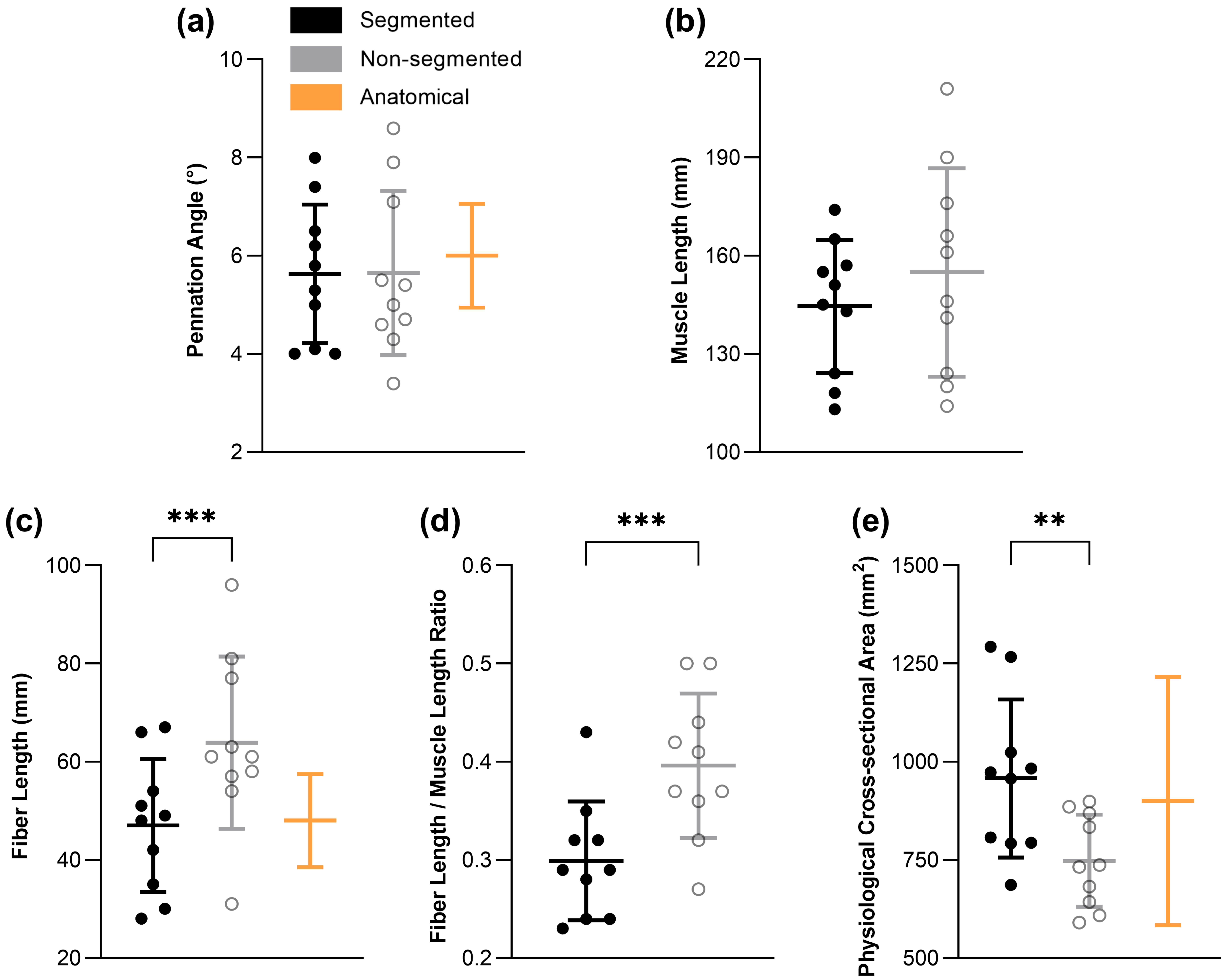}
    \label{fig:9}

\end{figure}

\noindent
\textbf{Fig. 9.}\hspace{15pt}Boxplots of segmented, non-segmented and anatomical
[2] (with permission) flexor digitorum superficialis (FDS) architectural
properties. Individual data from segmented and non-segmented architecture were
overlaid as scatter points (significance levels: * \textit{P} $<$ 0.05, **
\textit{P} $<$ 0.01, *** \textit{P} $<$ 0.001).

\begin{figure}[h]
    \centering
    \includegraphics[width=0.75\linewidth]{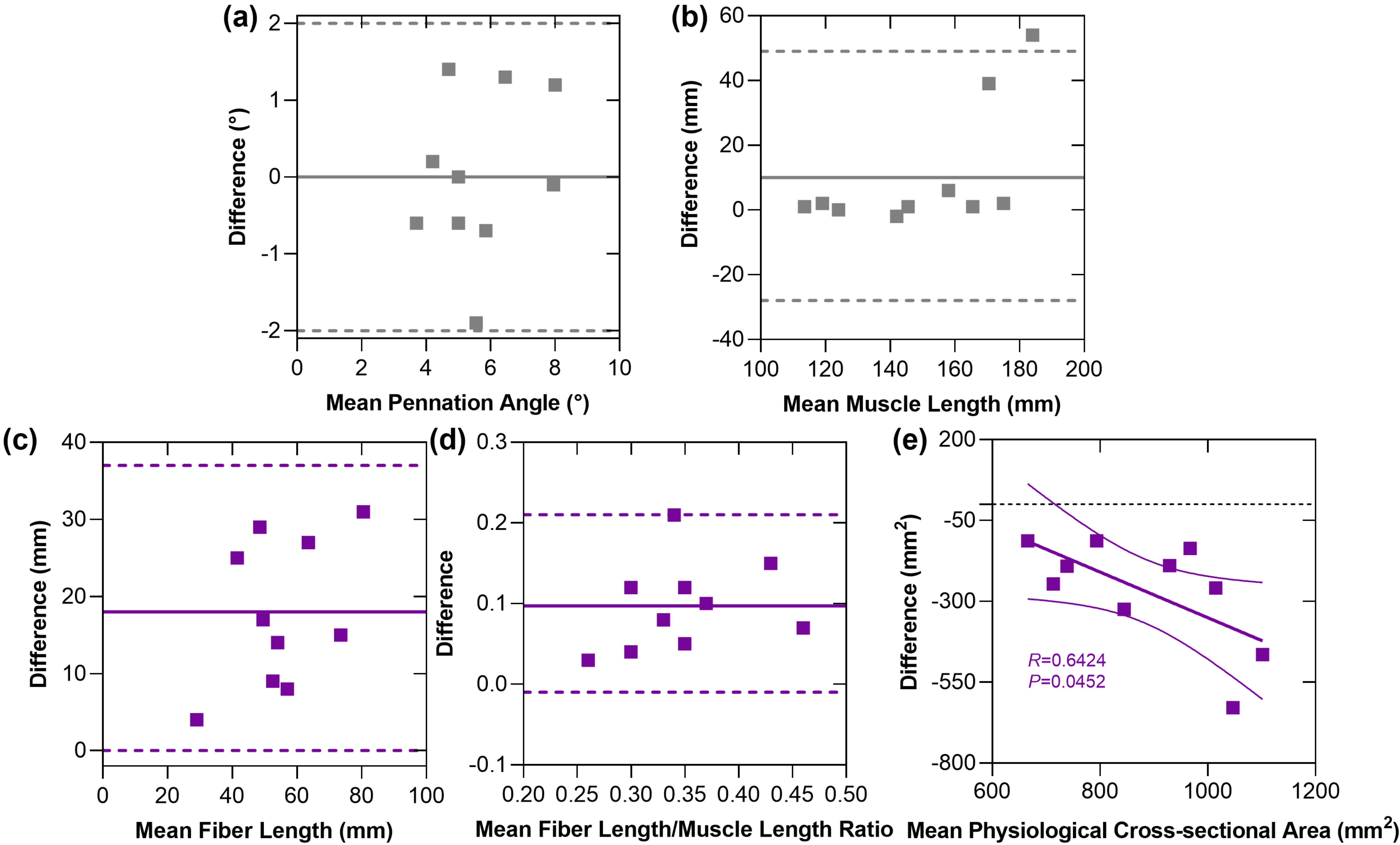}
    \label{fig:10}

\end{figure}

\noindent
\textbf{Fig. 10.}\hspace{15pt}Bland-Altman plots of the differences between
segmented and non-segmented FDS architectural properties. In (a)-(d), solid
lines: means, and dashed lines: 95\% confidence intervals. In (e), hyperbola:
95\% confidence limits. For (a) and (b), the differences represented by the grey
dots and lines were distributed around zero, while for (c)-(e) the significant
differences represented by purple in other properties were located on one side of
zero.

\newpage
\begin{figure}[h]
    \centering
    \includegraphics[width=0.75\linewidth]{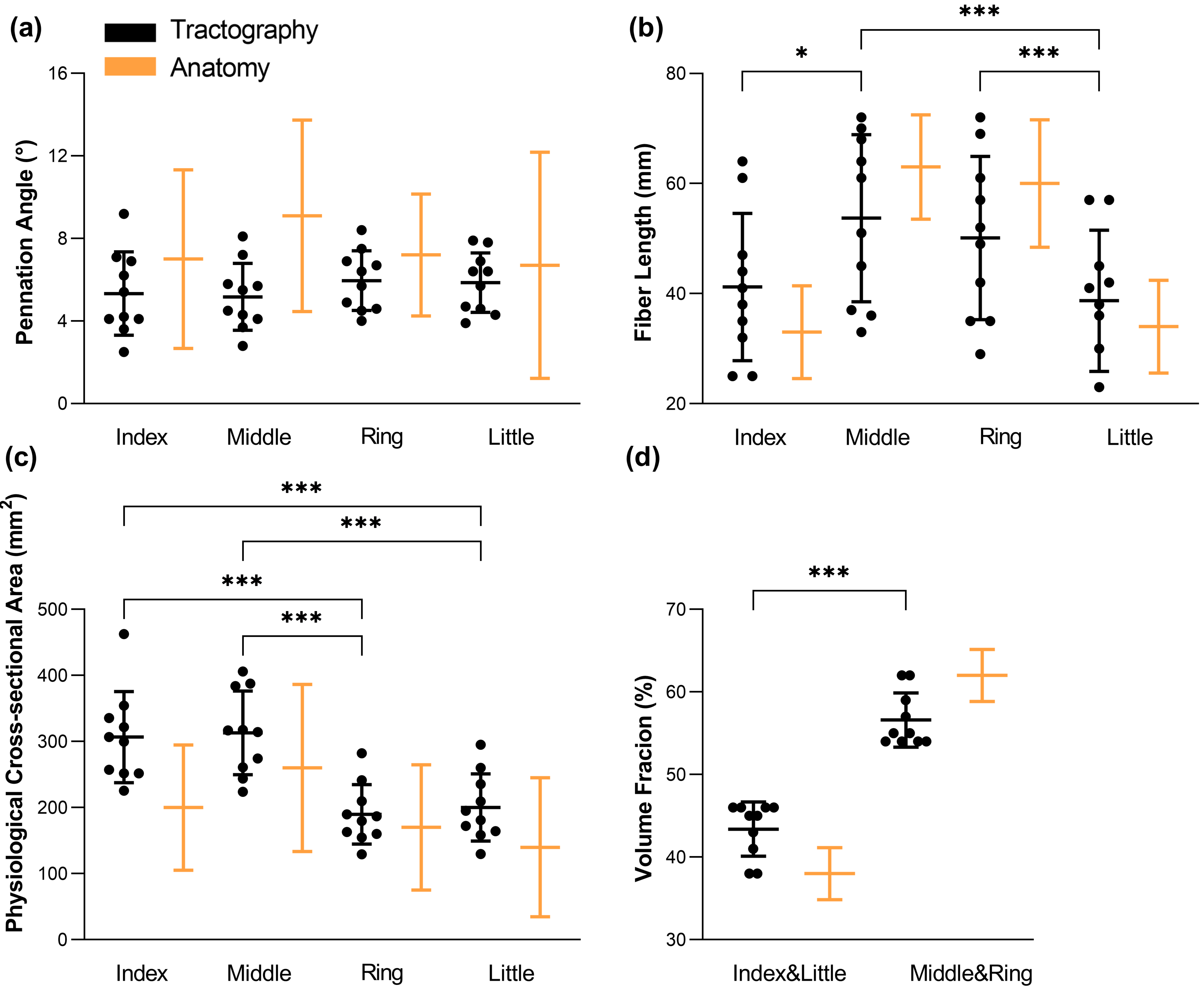}
    \label{fig:11}

\end{figure}

\noindent
\textbf{Fig. 11.}\hspace{15pt}Box plots of compartment architectural properties
from tractography and anatomy [1] (with permission). Individual data from
tractography-derived properties were added as scatter points (significance
levels: * \textit{P} $<$ 0.05, ** \textit{P} $<$ 0.01, *** \textit{P} $<$ 0.001).

\begin{figure}[h]
    \centering
    \includegraphics[width=1\linewidth]{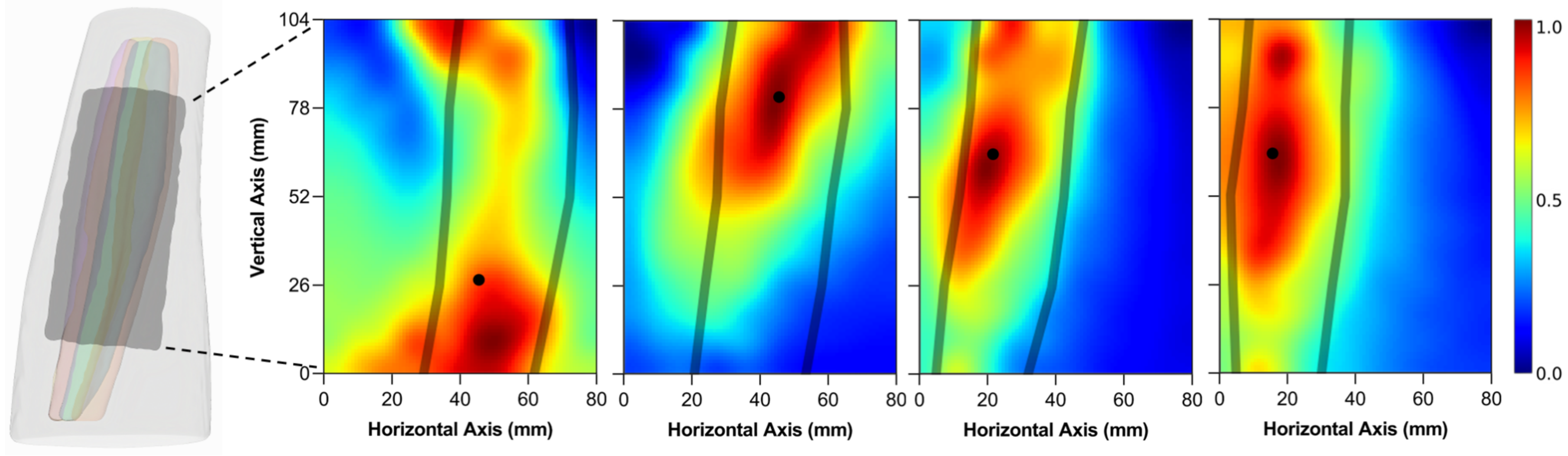}
    \label{fig:12}

\end{figure}

\noindent
\textbf{Fig. 12}.\hspace{15pt}Muscle activation maps (normalized RMS heatmaps),
superimposed with electromyogram centers (black dot) and projected boundaries of
muscle compartments (black curves), for index, middle, ring and little fingers
respectively. Horizontal and vertical axis displayed the actual distances
calculated based on the electrode size. The left illustration showed the muscle
compartments and the electrode surface in the MRI-derived forearm mask.

\newpage
\begin{figure}[h]
    \centering
    \includegraphics[width=0.63\linewidth]{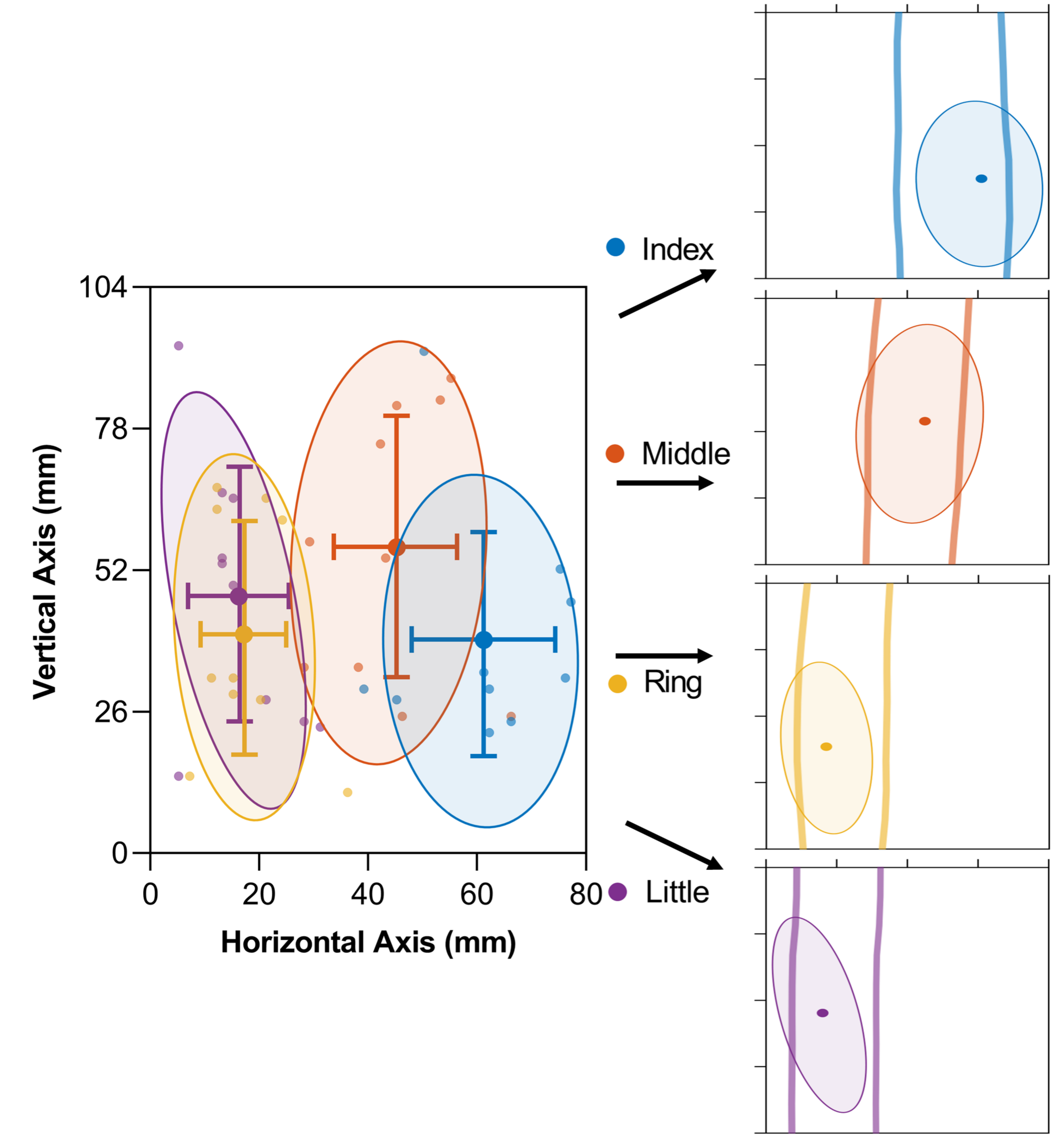}
    \label{fig:13}

\end{figure}

\noindent
\textbf{Fig. 13.}\hspace{15pt}Scatter plots, box plots and confidence ellipse
regions of the electromyogram centers of 10 subjects (left), and the confidence
regions superimposed with the inter-subject average boundaries of the muscle
compartments (right). The right side arranged the compartments longitudinally to
show the overlapping regions between them.

\end{document}